\documentstyle[prb,aps,amsfonts,amssymb,preprint, tighten]{revtex}
\begin{document}
\draft
\title{Peculiarities of anharmonic lattice dynamics and thermodynamics of 
alkaline-earth metals}
\author{M.I.Katsnelson}
\address{Institute of Metal Physics, Ekaterinburg 620219, Russia}
\author{A. V. Trefilov, M. N. Khlopkin, and K. Yu. Khromov}
\address{Russian Research Centre ``Kurchatov'' Institute, Moscow 123182, Russia}
\maketitle
\begin{abstract}
The calculations are performed for a broad range of the properties of Ca and Sr in the
fcc and bcc phases. A detailed information on the magnitude and character of
temperature dependence of anharmonic effects in the lattice dynamics over the
entire Brillouin zone (frequency shifts and phonon damping, Gruneisen
parameters) is given. A detailed comparison of the computational results for 
the heat
capacity and thermal expansion with the experimental data is carried out; the
theoretical results are in good agreement with the experiment.
\end{abstract}
\pacs{63.20-e, 63.20.Ry}

\narrowtext
\section{Introduction}

The investigation of anharmonic effects (AE) in lattice dynamics is a classical
problem of solid-state physics. It is important, particularly, because of the
role these effects can play in phenomena associated with structural phase
transitions and melting in crystals (see, e.g. \cite{1,2,3}). At the same time,
obtaining any information about the magnitude and scale of AEs from experiment
and theory is a difficult problem. The experimental study of such "basic" AEs
as the frequency shift and damping of phonons is very difficult and leads to a
large uncertainty in the results (see, e.g., the data presented in \cite{4,5}
for bcc and fcc metals, respectively). Up to now first-principles microscopic
calculations of AEs have been performed for one point of the Brillouin zone (N) in
the bcc phase of Zr and four points (N,P,$\omega$,G) in Mo \cite{6}.
Detailed information
about AEs in the entire Brillouin zone and their temperature dependence has
been obtained in \cite{4,7} on the basis of pseudopotential theory for the bcc
phases of alkali and alkaline-earth metals. For these metals the most striking
manifestations of AEs are due to the "soft-mode behavior"(the anomalous
temperature dependence of the phonon frequencies) of the $\Sigma_4$ branch.
It is of
interest to calculate AEs for the "general position", i.e.  for crystals which
do not possess soft vibrational modes. Such crystals include most metals with
close-packed structures, for example, fcc. In a recent paper \cite{8}
the results of
calculations performed for the anharmonic effects in lattice dynamics of Ir 
having the fcc structure have
been presented. In order to understand the specific features of AEs in lattice
dynamics, related to structural phase transitions, it is interesting to
investigate them in several phases for polymorphic metals. A classical example
of structural phase transitions in metals is the fcc-bcc temperature transition
in Ca and Sr \cite{9}. Therefore studies of AE features in these metals seem
to be important. In the present paper the AE features in Ca and Sr lattice 
dynamics
and thermodynamics have been investigated basing on microscopic calculations of
the heat capacity, thermal expansion, temperature frequency shifts, damping of
phonons and Gruneisen parameters.

\section{Approximations and computation procedure}
\label{sec:form}

As noted in Introduction, at present, {\it ab initio\/} calculations of
phonon frequency
with regard to AE have been only done for some highly symmetric points of the
Brillouin zone in a few metals \cite{6}. Consistent
{\it ab initio \/} calculations of
such
quantities as phonon damping, as well as thermal expansion coefficients and 
other integral
anharmonic effects in metals are now extremely difficult. Therefore in works
reported so far \cite{2,4,5,7,8} particular models of interatomic interactions
were
used. For example, for alkali metals \cite{2,4} and Ir \cite{8} the pair
interaction
approximation basing on pseudopotential theory was used, which gives a reliable
description of a variety of lattice properties of these metals. Construction of
a similar model for alkaline-earth metals appears to be difficult because of
the specific features of electron structure. The fcc phase of Ca and Sr has
many van Hove singularities in the density of electron states near the Fermi
level $E_F$ and, therefore, for the adequate description of lattice properties
allowance for related contributions to the total energy is important \cite{10}.
At
the same time, the Fermi surface for these metals is close to that in the
approximation of almost free electrons, and parameter
$|V_{\bbox{g}}|/E_F$ (where {\bf g} is the
reciprocal lattice vector,  $|V_{\bbox{g}}|$
is the Fourier component of pseudopotential)
is small (less than $0.1$) \cite{10}.
As shown in \cite{11}, under these conditions the
proximity of the Fermi surface to the Brillouin zone boundary essentially
contributes to the elastic moduli and to frequencies of long-wave phonons
with the wave vector
\begin{equation}
q \lesssim q_c = g_1 \sqrt {\frac{|V_{\bbox{g}}||E_c -E_F|}{E_F}} \label{1}
\end{equation}
where $E_c$ is the van-Hove singularity nearest to $E_F$, $g_1$  is the
minimum vector
of reciprocal lattice while the corresponding anomalous contributions to 
frequencies of
phonons with $q>q_c$  are much weaker. Therefore it may be expected
that for the
description of thermodynamics of fcc phase of Ca and Sr for
$T \gtrsim \Theta_D q_c/g_1 \approx
0.1\Theta_D$ (where $\Theta_D$ is the
Debye temperature) the ordinary approximation of pairwise interactions
corresponding to the second order to perturbation theory for
$|V_{\bbox{g}}|/E_F$ would be
applicable. This model was successfully used in \cite{7} for the
description of 
the phonon spectra of Ca and Sr in the high temperature bcc phase.

In the present paper the
phonon spectra, thermodynamic and anharmonic properties were calculated
using the model corresponding to the second order of perturbation theory for
pseudopotential (see, e.g. \cite{4,12}). For the latter the Animalou-Heine
expression
\widetext
\begin{equation}
V(q)=-\frac{4\pi Z e^2}{q^2}\left [\cos qr_0 +U\left ( \frac{\sin qr_0}
{qr_0}-\cos qr_0 \right)\right] \exp \left[-0.03\left(\frac{q}{2k_{F_0}}
\right )^4\right] \label{2}
\end{equation}
\narrowtext
was used, where $Z=2$ is the ion charge, $e$  is the electron charge, $r_0$,
U  are
the pseudopotential parameters listed in table \ref{table1},
$k_{F_0}$ is the Fermi momentum at zero
pressure. For the screening the Geldart-Taylor approximation \cite{13} with the
Ceperley-Alder expression for correlation energy was used (see \cite{12}  for
details).  To eliminate the influence of fcc band structure peculiarities 
(the closeness of the Fermi surface to the bounadaries of the Brillouin
zone), parameters $r_0$
and
$U$ were fitted to the properties of high-temperature bcc phase:
$\Omega=\Omega_{exp}(T=T_s)$ and $\Theta_D=\Theta_D^{exp}(T=T_s)$ for Sr
and $\Omega=\Omega_{exp}(T=T_s)$ and $C'=C'_{exp}$
for Ca. Here $T_s$ is the temperature of bcc-fcc transition
(see table \ref{table1}),
index "exp" indicates the related experimental value,
$C'=1/2(C_{11}-C_{12})$ is
one of the shear moduli.

For the AE calculations the standard anharmonic
perturbation theory was used \cite{14,4}.
With an accuracy up to terms of $\varkappa^2$ order,
where $\varkappa=(m/M)^{1/4}$ is the adiabatic parameter,
$m$, $M$ are the masses of electron
and ion, respectively, the Hamiltonian of phonon subsystem can be written as:
\begin{equation}
H=H_0+H_{qh}+H_3+H_4 \label{3}
\end{equation}
where
\begin{equation}
H_0=\sum_\lambda \omega_\lambda b_\lambda^+b_\lambda  \label{4}
\end{equation}
is the Hamiltonian of harmonic approximation,
$\lambda\equiv\bbox{q}\xi$, \bbox{q}
is the wave vector of
phonon, $\xi$ is the number of phonon branch, $b_\lambda^+$,
$b_\lambda$
are the phonon operators of creation
and annihilation.
\begin{equation}
H_{qh}=\sum_{\lambda i} \left (\frac{\partial \omega_\lambda}{\partial u_i}
\right )_{u_i=0} u_i b_\lambda^+b_\lambda      \label{5}
\end{equation}
is the quasiharmonic Hamiltonian, $u_i$ are the deformation parameters
(in cubic
crystals it is sufficient to allow for dilatation $u_1$,
$du_1 = d\ln\Omega$, where $\Omega$ is
the lattice volume)
\begin{eqnarray}
H_3&=&\sum_{\lambda_1\lambda_2\lambda_3}
V^{(3)} (\lambda_1, \lambda_2, \lambda_3)
Q_{\lambda_1} Q_{\lambda_2} Q_{\lambda_3}\nonumber \\
&=&\sum_{\lambda_1\lambda_2\lambda_3} \Phi^{(3)}
(\lambda_1, \lambda_2, \lambda_3) A_{\lambda_1} A_{\lambda_2} A_{\lambda_3}
\label{6}
\end{eqnarray}
\begin{eqnarray}
H_4&=&\sum_{\lambda_1\lambda_2\lambda_3\lambda_4} V^{(4)}
(\lambda_1, \lambda_2, \lambda_3, \lambda_4)
Q_{\lambda_1} Q_{\lambda_2} Q_{\lambda_3} Q_{\lambda_4} \nonumber \\
&=&\sum_{\lambda_1\lambda_2\lambda_3\lambda_4} \Phi^{(4)}
(\lambda_1, \lambda_2, \lambda_3, \lambda_4)
A_{\lambda_1} A_{\lambda_2} A_{\lambda_3} A_{\lambda_4} \label{7}
\end{eqnarray}
are the Hamiltonians of three- and four- phonon processes, respectively,
\begin{equation}
Q_\lambda=\frac{1}{\sqrt{2M\omega_\lambda}}A_\lambda,\quad A_\lambda=
b_\lambda+b_{-\lambda}^+ \label{8}
\end{equation}
is the operator of phonon coordinate, $-\lambda\equiv  - {\bf q},\xi$.
Here and below we put the
Planck constant as $\hbar = 1$. Expressions for the amplitudes of three- and
four-phonon processes $\Phi^{(3)}$, $\Phi^{(4)}$, in terms of the pseudopotential model used
are shown in \cite{4}. With an accuracy up to terms of $\varkappa^2$ order we have the
following expressions for anharmonic shifts of phonon frequencies  and damping
(see Appendix):
\begin{equation}
\Delta_\lambda=
\Delta_\lambda^{(qh)}+\Delta_\lambda^{(3)}+\Delta_\lambda^{(4)} \label{9}
\end{equation}
\begin{equation}
\Delta_\lambda^{(qh)}=-\gamma_\lambda \frac{\Delta\Omega}{\Omega}
\label{10}
\end{equation}
\widetext
\begin{eqnarray}
\Delta_\lambda^{(3)}&=&-18\sum_{\lambda_2\lambda_3} \left |
\Phi^{(3)}(\lambda, -\lambda_2, -\lambda_3)
\right |^2 \left \{(1+N_{\lambda_2}+N_{\lambda_3})\left[\frac{1}{\omega_\lambda+
\omega_{\lambda_2}+\omega_{\lambda_3}}\right.\right. \nonumber \\
&&+\left.\frac{\cal{P}}{\omega_{\lambda_2}+\omega_{\lambda_3}-
\omega_\lambda}\right]+
(N_{\lambda_2}-N_{\lambda_3})\left. \left[\frac{\cal{P}}{\omega_{\lambda_3}-\omega_{\lambda_2}-
\omega_\lambda} - \frac{\cal{P}}{\omega_{\lambda_2}-\omega_{\lambda_3}-
\omega_\lambda} \right] \right \} \label{11}
\end{eqnarray}
\begin{equation}
\Delta_\lambda^{(4)}=12\sum_{\lambda'} \Phi^{(4)}(-\lambda,\lambda,\lambda',
-\lambda')(1+2N_{\lambda'}) \label{12}
\end{equation}
\begin{eqnarray}
\Gamma_\lambda&=&18\pi\sum_{\lambda_2\lambda_3} \left | \Phi^{(3)}(\lambda,
-\lambda_2, -\lambda_3)
\right |^2 \left \{(1+N_{\lambda_2}+N_{\lambda_3})\delta(
\omega_{\lambda_2}+\omega_{\lambda_3}-\omega_\lambda)\right. \nonumber \\
&&+(N_{\lambda_2}-N_{\lambda_3})\left. \left[\delta(\omega_{\lambda_3}-\omega_{\lambda_2}-
\omega_\lambda) - \delta(\omega_{\lambda_2}-\omega_{\lambda_3}-
\omega_\lambda) \right] \right \} \label{13}
\end{eqnarray}
\narrowtext
here
\begin{equation}
\gamma_\lambda=-\frac{\partial \ln \omega_\lambda}{\partial \ln \Omega}
\label{14}
\end{equation}
are the Gruneisen parameters, $\Delta\Omega$
is the change in the crystal volume due to the thermal expansion
\begin{equation}
N_\lambda=\frac{1}{\exp(\omega_\lambda/T)-1} \label{15}
\end{equation}
is the Planck
distribution function, $\cal{P}$ is the symbol of principle value. In the particular
calculations the approximation
\begin{equation}
{\cal{P}} \frac{1}{x} \approx \frac{x}{x^2+
\varepsilon^2}
\end{equation}
\begin{equation}
\delta(x)\approx\frac{1}{\pi}\frac{\varepsilon}{x^2+\varepsilon^2} \label{16}
\end{equation}
was used for different small positive
$\varepsilon$ with subsequent extrapolation $\varepsilon\to 0$. The details of
the calculations, in
particular,  the summing up over the Brillouin zone is discussed in detail in
\cite{4}.

\section{Calculations of phonon spectra in harmonic approximation}

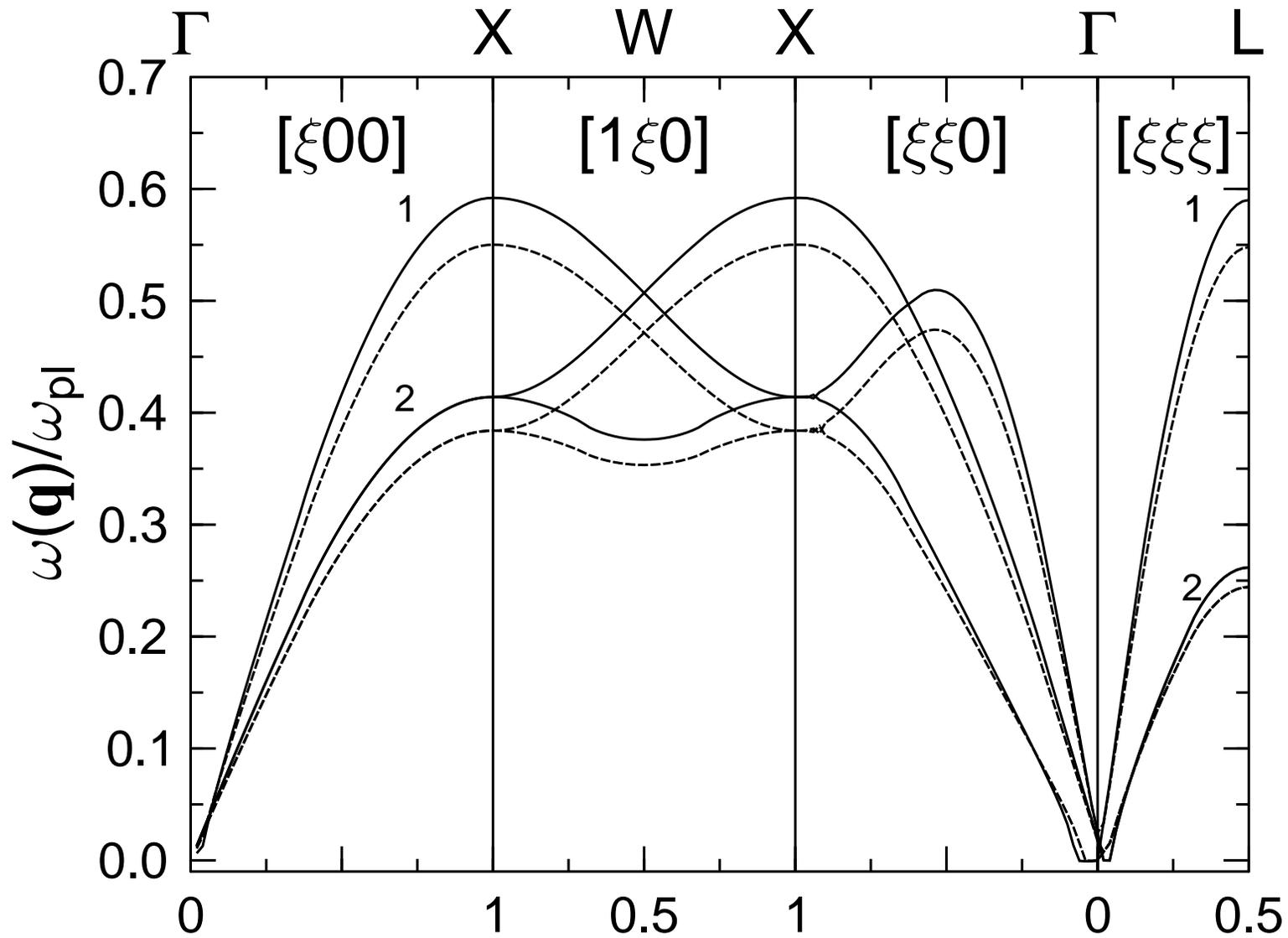
\begin{figure}
\caption{Phonon spectrum of fcc phase of Ca (solid line) and Sr (dashed line)
in the harmonic approximation in units of ion plasma frequency
$\omega_{pl}=(4\pi Z e^2/M\Omega)^{1/2}$.
Indices 1 and 2 denote longitudinal and transverse branches respectively.}
\label{fig1}
\end{figure}

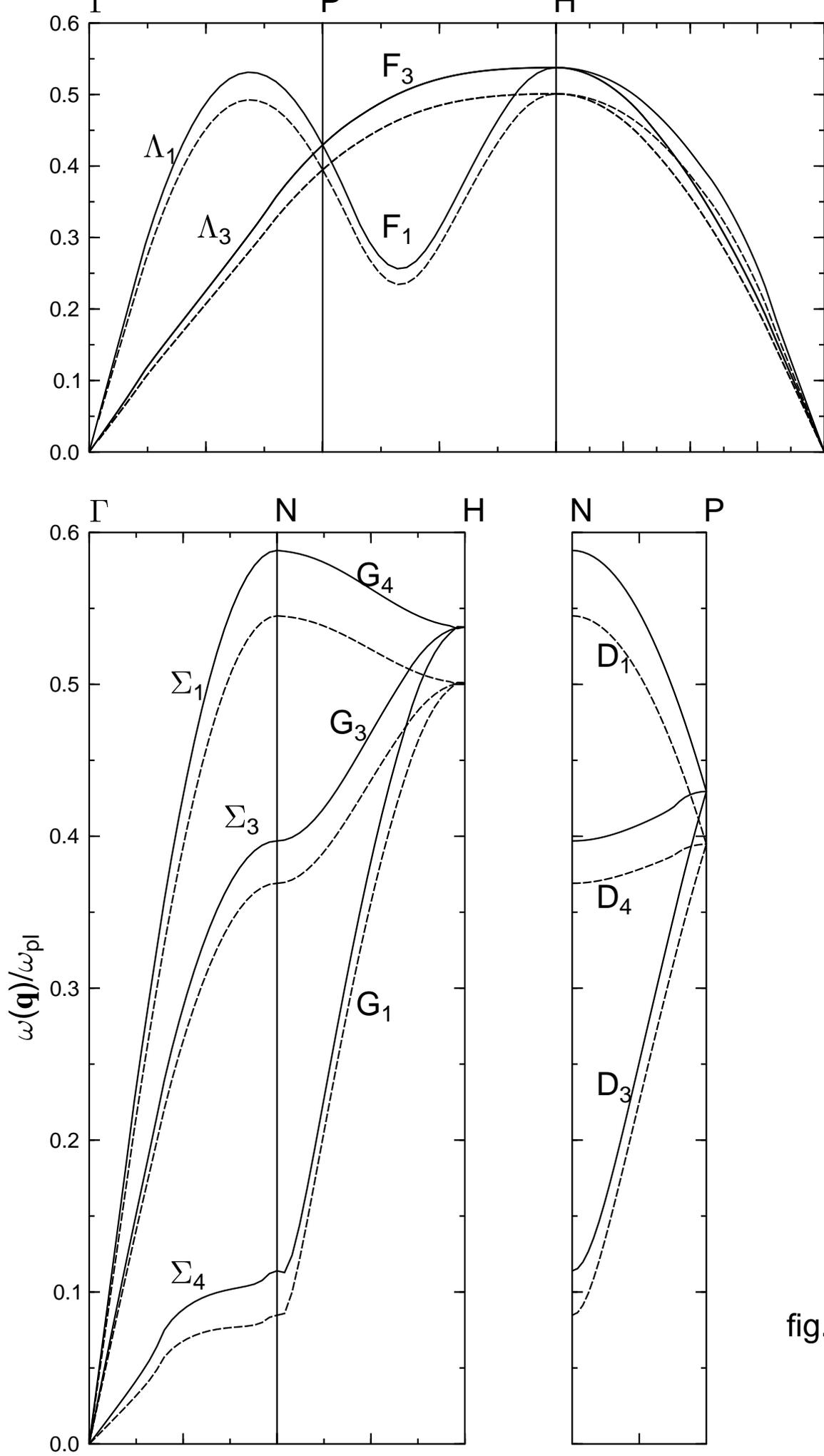
\begin{figure}
\caption{Phonon spectra for bcc phase of Ca (solid line) and Sr (dashed line)
in units  $\omega_{pl}$ in the harmonic approximation.}
\label{fig2a}
\end{figure}

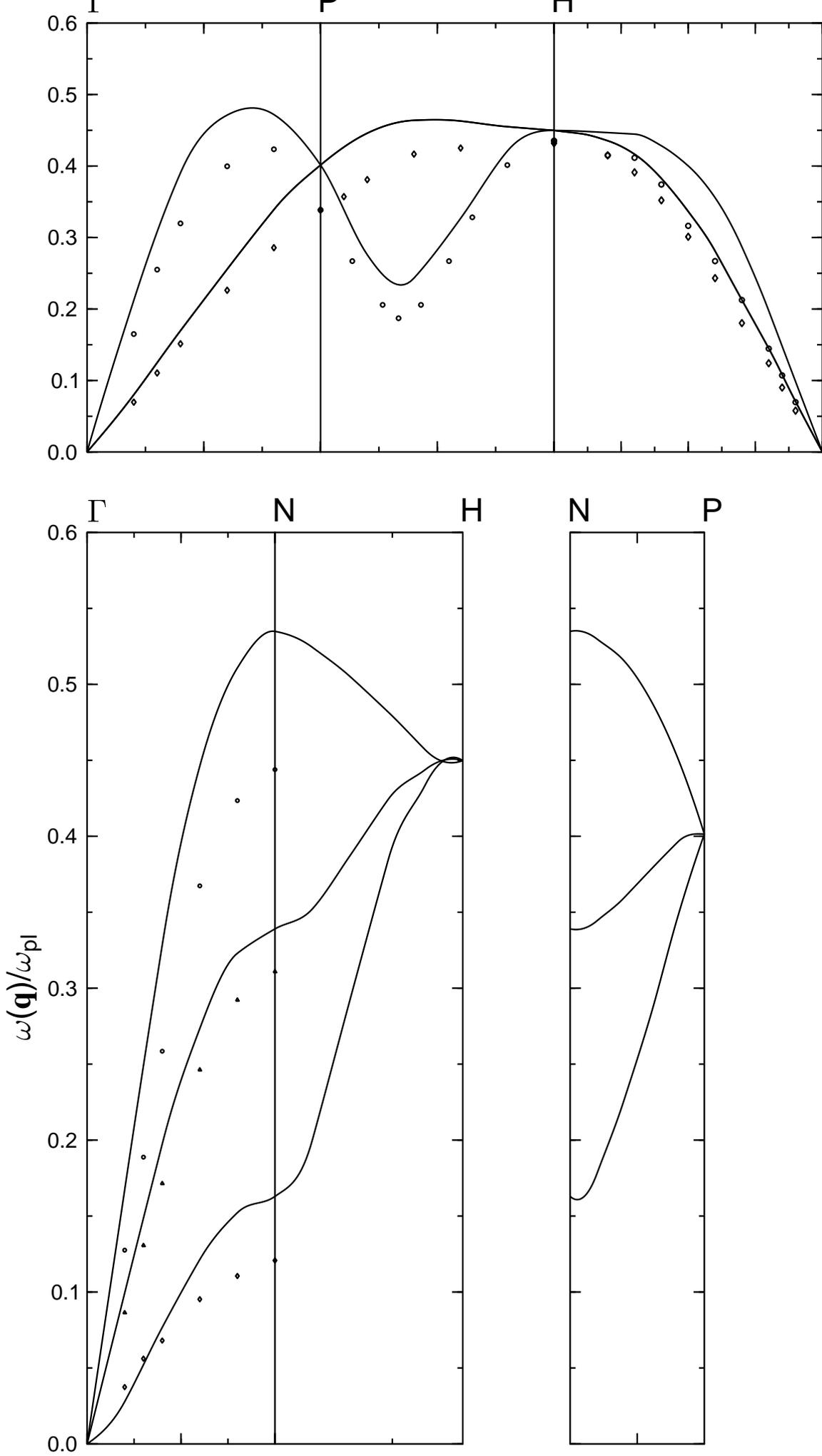
\begin{figure}
\caption{Phonon spectra for bcc phase of Sr at $T = T_s$;
solid line is the calculation (with the account for anharmonic shift, see
below), points - the experimental data from \cite{exp1}.}
\label{fig2b}
\end{figure}

The results of calculations of phonon spectra and Gruneisen parameters
(\ref{14}) for
fcc and bcc phases of Ca and Sr are shown in Figs. \ref{fig1}-\ref{fig4}.
As noted above, the
theoretical model used is unapplicable for the description of phonon spectra in
the fcc phase of these metals in the nearest vicinity of point $\Gamma$
because of
unaccounted effects of Fermi surface proximity to the Brillouin zone boundaries
\cite{10}. It follows from Fig. \ref{fig1} that $q_c$c from Eq. (\ref{1})
is of the
order of $0.1 g_1$. Fig. \ref{fig2a}
shows that the phonon spectrum in the bcc phase of Ca and Sr is qualitatively
similar to phonon spectra in the bcc phase of alkali metals \cite{4}, exhibiting the
same characteristic features: presence of "soft" $\Sigma_4$ branch and
prominent minimum
for branch F$_1$. In the fcc phase of these metals there are no soft mode
anomalies. Fig. \ref{fig2b} displays such phonon spectra of bcc Sr
for $T = T_s$ with
allowance for anharmonic frequency shifts and corresponding experimental data
in accordance with \cite{exp1}. It is seen that there is a reasonable agreement
between the calculated phonon spectra and experimental data. Unfortunately, any
direct experimental data for  $\omega({\bf q})$  in the fcc phase of Ca and Sr are
unavailable. The comparison of theory with the experiment for this phase will
be done in the next Section basing on the data on temperature dependence of
the lattice heat capacity. As seen from Figs. \ref{fig3}, \ref{fig4} the Gruneisen
model $\gamma_\lambda=$const is
absolutely unapplicable, even qualitatively, for the description of bcc phases
in alkali metals. In the fcc phase the {\bf q}-dependence is much weaker (provided
that the vicinity of point $\Gamma$, where the model itself becomes unapplicable, is
not considered). This seems to be because of lack of "soft modes" in the phonon
spectra of fcc phase.

\section{Anharmonic effects in lattice dynamics}

The calculation results for the  anharmonic effects in the lattice dynamics
for Ca and Sr
are given in Figs. \ref{fig5a}-\ref{fig9b}. The temperature dependencies of
the frequency shift and
phonon damping are shown for Ca as an example (Figs. \ref{fig5a}, \ref{fig5b},
\ref{fig8a}, \ref{fig8b});
for Sr the anharmonic
effects are somewhat stronger than for Ca, nevertheless all dependencies are
similar.

Figs. \ref{fig5a}, \ref{fig5b} displays temperature dependencies of phonon
frequency shifts:
\begin{eqnarray}
\Delta\omega(T)&=&\omega(T)-\omega(T=0) \nonumber \\
&=&
\Delta_\lambda^{(qh*)}+\Delta_\lambda^{(3*)}+\Delta_\lambda^{(4*)}
\label{extra1}
\end{eqnarray}
Here $\Delta_\lambda^{(qh*)}$, $\Delta_\lambda^{(3*)}$,
$\Delta_\lambda^{(4*)}$ are values determined by Eqs.
(\ref{10}-\ref{13}) respectively without the contribution of zero point 
oscillations, since such contributions enter both $\omega(T)$ and
$\omega(T=0)$ and are compensated in the difference (\ref{extra1}).
Note also that since fitting of the pseudopotential papameters was done
at the volume $\Omega=\Omega_{exp}(T=T_s)$ (see Section \ref{sec:form}),
$\Delta\Omega$ entering right hand side of Eq. (\ref{10}) is
determined by the following relation: $\Delta\Omega=\Omega(T)-\Omega(T_s)$

It is seen that both in the bcc and in fcc phases $\Delta\omega(T)$  early
(at  $T\approx 0.1\div 0.2 T_{pl}$) comes to
high-temperature behavior $\Delta\omega(T)\sim T$. In the bcc
phases of Ca and Sr
the typical
soft-mode behavior $d\omega/dT > 0$
for $\Sigma_4$ branch takes place (see the data for $N_4$ in Fig.
\ref{fig5b}). As pointed out in \cite{7} the anharmonic effects in the
bcc phases of
alkaline-earth and alkaline metals are qualitatively similar in the whole, but,
however, the scale of these effects in alkaline-earth metals is essentially
larger. In the fcc phase of Ca and Sr there are no "soft modes" and
$d\omega/dT < 0$
for all vibrational modes.

Figs. \ref{fig6a}, \ref{fig6b} shows the {\bf q}-dependence  of the relative
frequency
shift in Ca and Sr. The comparison of the results for the fcc phase
(Fig. \ref{fig6a})
with those obtained in \cite{8} for Ir indicates that the positions of frequency
shift minima and maxima in the Brillouin zone depends essentially on the ion
charge; in Ir at $Z = 4.5$ maxima of
$\delta$ are reached at symmetric points X and L,
while in alkaline-earth metals they are significantly shifted. Moreover, in
contrast with the temperature dependencies $\delta(T)$, there is a noticeable
difference in
the behavior of $\delta(\bbox{q})$ in Ca and Sr
(cf., for example, the dependence in $\Gamma-L$
direction in Fig. \ref{fig6a}).

For the bcc phases of Ca and, particularly, Sr, $\delta(\mathbf{q})$
has a prominent maximum in the narrow region of $\bf q$-space near point N for
the soft
branch $\Sigma_4$ for which $\delta>0$ (see Fig. \ref{fig6b}).

On example of Ca, Figs. \ref{fig7a}, \ref{fig7b} show the
contributions of three- and four- phonon processes to the frequency shift. In
both phases these contributions have opposite signs in the larger part of
the Brillouin zone  ($\Delta_3<0, \Delta_4>0$)
The mentioned above "soft mode" behavior of the branch $\Sigma_4$  in
the bcc phase is due to a sharp increase in $\Delta_4$  as in alkali metals
\cite{4} and in
$BaTiO_3$-type ferroelectrics \cite{1}.

In Figs. \ref{fig8a}, \ref{fig8b} the temperature dependence of phonon
damping in the symmetric points of Brillouin zone is shown for Ca as an
example.
First of all it should be pointed out that, in accordance with Eq. (\ref{13})
the damping
does not vanish even at $T=0$ . It is due to three-phonon processes
(phonon decay into two phonons)
and has a smallness of $\varkappa^2$ order. It is seen
from Figs. \ref{fig8a}, \ref{fig8b}
that for Ca it is of the order of $0.2\div0.5$\% of phonon frequency. As
in the case of anharmonic frequency shift, it is seen that transition to
asymptotic $\Gamma\sim T$ is reached very early (at $T\approx 0.1T_{pl}$).

Figs. \ref{fig9a}, \ref{fig9b} show the
{\bf q}-dependence of relative damping of phonons $\eta$ in Ca and Sr.
In the fcc phase $\eta$
has essentially nonmonotonous dependence on the wave vector, reaches values of
about 25\% in Sr and does not exceed 12\% in Ca.
In the bcc phase of Sr in the vicinity
of point N 
$\eta$ reaches about 1 for the soft mode $\Sigma_4$, which formally
indicates inapplicability of the anharmonic perturbation theory in this narrow
region. At the same time in the larger part of the Brillouin zone the
anharmonic perturbation theory seems to be applicable. In view of this a
problem arises on accurate separation of the contribution of the vicinity of
point N to integral anharmonic characteristics such as free energy. This
question will be considered elsewhere \cite{1prim}.

\section{Analysis of experimental data on heat capacity and thermal expansion}

\subsection{Heat capacity}

The lattice heat capacity is determined by the
phonon state density $g(\omega)$. Some characteristics of the latter, which can be
used for comparing our description of phonon spectra with the experiment can
be found from the analysis of temperature dependence of heat capacity at the
constant pressure $C_p(T)$.

Here we carried out the separation of the lattice
components from the heat capacity measured in the experiment and determined
some average frequencies (momenta) of phonon spectrum, without using any
model.

In our analysis we used the following set of data: for Sr -- direct
experimental data from \cite{boerio}, for Ca -- the data from handbooks;
in the region
2-50 K -- the data from handbook \cite{hultgren} which in the region 2-10 K
is based on \cite{roberts},
and in the region 50-600 -- the data taken from handbook \cite{landolt}.

In the analysis we
neglected the temperature dependence of the electron heat capacity coefficient
and assumed that the anharmonism is weak. The assumption on smallness of
anharmonic  contributions to the thermodynamical properties of metals is
justified as it was checked by a direct calculation and found to be true even
for alkali metals having soft phonon modes \cite{2}. Under these assumptions the
heat capacity measured experimentally at a constant pressure, $C_p$ is described
by the relations:
\begin{equation}
C_p = C_{ph} + C_a   \label{ex1}
\end{equation}
\begin{equation}
C_a=\left\{\gamma + ( A - \gamma)(C_{ph}/3R)^2\right\} T \label{ex2}
\end{equation}
where

$C_{ph}$ is the phonon component of the heat
capacity in the harmonic approximation,

$C_a$ is the sum of contributions to the
heat capacity, having the linear temperature dependence and coming from the
anharmonic effects, thermal expansion of the lattice and conduction electrons;

$\gamma$ is the coefficient of electron heat capacity at low temperatures,

$A$ is the
coefficient of linear in temperature term at high temperatures,

$R$ is the gas
constant.

The interpolation formula (\ref{ex2}) for $C_a$ gives the corresponding linear
temperature asymptotics both at low- and high-temperatures and ensures a smooth
transition between the low and high temperature asymptotics in accordance with
the law similar to Nernst-Linderman formula
\cite{reznitskii}.

In the high-temperature
region the phonon component of heat capacity was described by the expression
proposed by Naumov \cite{naumov}
\begin{equation}
C_{ph}=3R\left\{1-\frac{1}{12}\left(\frac{\Omega_2}{T}\right)^2+
\frac{1}{240}\left(\frac{\Omega_4}{T}\right)^4 +
\varphi\left(\frac{\Omega_*}{T}\right)\right\} \label{ex6}
\end{equation}
\begin{equation}
\varphi(z)=\frac{z^2exp(z)}{\left(1-exp(z)\right)^2}-
\left(1-\frac{1}{12}z^2+ \frac{1}{240}z^4\right) \label{ex7}
\end{equation}
Here the asymptotic expansion
of phonon heat capacity over small parameter $z = \Omega/T$  is used:
\begin{equation}
C_{ph}=3R\left(1-\sum_{n=2}^{\infty}\frac{(n-1)B_n}{n!}\left
(\frac{\Omega_{n}}{T}\right)^{n}\right) \label{ex8}
\end{equation}
where $B_n$ are the
Bernulli numbers ($B_2 = 1/6$, $B_4 = -1/30$, $B_66 = 1/42$,
$B_8 = - 1/30$, $B_{10} = 5/66$ etc.
For the odd n, beginning from n = 3, all $B_n$  = 0).

In (\ref{ex6}) the second and
fourth order correction for $\Omega/T$ are separately written, and function
$\varphi (\frac{\Omega_*}{T})$
allows for all higher corrections in the " Einstein" approximation, i.e.
assuming $\Omega_{n} = \Omega_{*}$ for all $n \ge 6$.
The values $\Omega _{n}$ characterize momenta (mean
frequencies) of phonon spectrum according to the relation:
\begin{equation}
(\Omega_{n})^{n}=<\omega^n>=\left.\int_0^{\infty}g(\omega)\,
\omega^{n}\,d{\omega}\right/\int_0^{\infty}g(\omega)\,d{\omega} \label{ex9}
\end{equation}

The value $\gamma$
entering $C_a$ (\ref{ex2}) was determined in the standard manner: by fitting of
heat capacity in the low temperature region (particularly, 2-5 K for Ca and
5-15 K for Sr) by the relation $C_p=\gamma T+\beta T^3$.
The values $\gamma$, $\beta$ as well as
the limiting low-temperature value of the Debye temperature $\Theta_{LT}$
related to $\beta$
by relation $\beta=12\pi^4R/(5\Theta^3_{LT})$
are listed in Table \ref{table2}. The estimations of $\Theta_{LT}$ we
obtained from the set of data used reasonably agree with those available in the
literature (see \cite{boerio,roberts}).

The value $A$ together with  $\Omega_2$ , $\Omega_4$ and $\Omega_*$
were determined using the least
square method by the fitting of heat capacity by relations
(\ref{ex6}, \ref{ex7})
in the temperature region 40-600 K for Ca and 32-350 K for Sr.
Within this temperature range these relations described the
experimental results with least square deviation of the order of 1\% for
Ca and 0.3\% for Sr.
The parameters $A$, $\Omega_2$, $\Omega_4$  and $\Omega_*$ determined by the
least square method are shown in Table \ref{table2}. The table also gives the
limiting high-temperature value of Debye temperature $\Theta_{HT}=\Theta_D(T)$
where $\Theta_{LT} \ge T < T_m$
($T_m$ is the melting temperature) related to the second
momentum of the phonon spectrum by:
$\Omega_2 = \Theta_{HT} \sqrt{3/5}$ \cite{maradudin}.

It should be
pointed out that the lower value of the least square deviation for Sr
is due to that over the whole temperature range the experimental data
from the same source were used. On the other hand, its higher value for
Ca seems to be due to non-smooth joining of the data at the transition from
one source to another one.

The analysis performed permitted the
contribution of $C_{ph}$ in the harmonic approximation to be separated from
the total heat capacity and the electron and anharmonic contributions
to be left out. Some phonon spectrum momenta are expressed directly
via phonon heat capacity integrals \cite{junod}.
\begin{equation}
<\omega>=2\int_0^{\infty}\left(1-\frac{C_{ph}}{3R}\right)\,dT
\end{equation}
\begin{equation}
<\omega^{-1}>=\frac{3}{\pi^2}\int_0^{\infty}\frac{C_{ph}}{3R}\,T^{-2}\,dT
\end{equation}
\begin{equation}
<\omega^{-2}>=0.138651\int_0^{\infty}\frac{C_{ph}}{3R}\,T^{-3}\,dT
\label{ex10}
\end{equation}
\begin{equation}
<\omega^{-1}\,log\,\omega>=\frac{3}{\pi^2}\int_0^{\infty}\frac{C_{ph}}
{3R}\,log\left(\frac{T}{0.70702}\right)\,T^{-2}\,dT
\end{equation}

We calculated these momenta with the integration over the experimental
points in the temperature region 10-300 K and outside this region -- over
low temperature asymptotic $C_{ph}=\beta T^3$  and high temperature asymptotic
in the Debye spectrum model as in \cite{mirmelstein}. The mean
frequencies corresponding to these momenta are listed in Table \ref{table2}. The
meanings of $\Omega_{-2}$, $\Omega_{-1}$, $\Omega_1$
corresponding to that of  $\Omega_n$ in relation \ref{ex9}
for n = -2, - 1 and 1  while the meaning of
$\Omega_{log}$ is determined by
relation:
\begin{eqnarray}
\log(\Omega_{log})&=&\frac{<\omega^{-1}\,\log\omega>}
{<\omega^{-1}>}\nonumber \\
&=&\int\limits_0^{\infty}\frac{g(\omega)\log{\omega}}{\omega}\,d
{\omega}\Bigg / \int\limits_0^{\infty}\frac{g(\omega)}{\omega}\,d{\omega}
\end{eqnarray}
The contribution of $C_{ph}$ in the harmonic approximation separated from the total heat
capacity is shown  in Figs. \ref{fig10a}, \ref{fig10b}.

\subsection{Thermal expansion}

The coefficient of thermal expansion $\alpha(T)$ contains two contributions:
the electron one
$\alpha_e$ and the lattice one $\alpha_l$.
In the low temperature region the electron contribution is
linearly dependent on temperature, while the lattice contribution is
proportional to $T^3$. This makes it possible to separate these contributions
fitting the experimental data by the dependence:
\begin{equation}
\alpha(T)=DT+ET^3
\end{equation}
In the paper \cite{white} the electron contribution to thermal expansion of Ca
and Sr
is determined. For the determination of the lattice contribution to thermal
expansion we used the following set of experimental data: in the low
temperature region --- the data from \cite{white} and in the high temperature
region ---
the data from \cite{novikova} for Ca and \cite{physval} for Sr. The phonon
contribution
to thermal expansion was determined as the difference of the experimental data
and the electron contribution. It was assumed that the electron contribution is
linear by temperature over the whole temperature region, its value being
determined by low temperature asymptotic in \cite{white}.
The lattice contribution to
thermal expansion separated from the experimental data is shown in
Figs. \ref{fig11a}, \ref{fig11b}.

The lattice contributions to the heat capacity and the coefficient of thermal
expansion, separated from the experimental data were used for the determination
of the Gruneisen lattice parameter by the formula:
\begin{equation}
     G=\frac{3 \alpha B V_m}{C_{ph}}
\end{equation}
where $B$ is the bulk modulus,
$\alpha$ is the temperature coefficient of linear expansion,
$V_m$ is the molar volume. The temperature dependence of the Gruneisen parameter
$\gamma(T)$ is shown in Figs. \ref{fig11a}, \ref{fig11b}.

\section{Calculations of heat capacity and thermal expansion}

Figs. \ref{fig10a}, \ref{fig10b} give the computation results of lattice heat 
capacity of Ca and Sr in
the harmonic approximation and the effective Debye temperature. It is seen
that, with the exception of the region of lowest temperatures where the phonon
spectrum model itself becomes inapplicable (see Section \ref{sec:form}),
the theory is  in
a perfect agreement with the experiment. It is especially well seen in
Table \ref{table2},
where the computation data are compared with the momenta of phonon density of
states, found from the experimental data.

Figs.  \ref{fig11a}, \ref{fig11b} displays the computation results
for the temperature dependence of the volume
$\Delta\Omega(T)$ and of the Gruneisen
parameter $\gamma(T)$ for Ca and Sr.It is related to the  Gruneisen
parameters by the following expression:
\begin{equation}
\gamma(T)=\frac{\displaystyle \sum_\lambda \gamma_\lambda N_\lambda
(1+N_\lambda)\omega_\lambda^2}{\displaystyle \sum_\lambda N_\lambda
(1+N_\lambda)\omega_\lambda^2} \label{eq61}
\end{equation}
One can see that the temperature dependence of the macroscopic
Gruneisen parameter is a rather smooth. At the same time,
our calculations show a strong temperature dependence of microscopic
Gruneisen parameters for the soft branch in BCC phases of Ca and
Sr (see Table \ref{table3}). It would be very interesting to
check this prediction experimentally.
 
\section{Conclusions}

In the present paper a simple, and at the same time, adequate
model of the interatomic interactions is proposed, which describes with a 
reasonably high
accuracy the lattice properties of fcc and bcc phases of Ca and Sr.  
As a result, the dependence of AE in the dynamics of lattice on its
geometry could be investigated. The comparison of the results obtained for the
fcc phase
of Ca and Sr with earlier results for Ir \cite{8} demonstrates an essential
dependence of AE on the character of itneratomic interactions for the given
crystal structure. As the question on the AE scale, particularly, for metals
with polymorphic transformation (in this case bcc-fcc) is one of the key
problems in the lattice dynamics, performance of further experimental
investigations is desirable.

\appendix

\section*{}

Here we shall give a short derivation of equations for the
temperature shift of frequencies and phonon damping (\ref{9}-\ref{13}).
In contrast to the
basing work \cite{14} and subsequent publications, where the diagram method for
Matsubara Green functions was used, we employ the method of two time Green
functions \cite{A1}. We shall begin with the chain of equations of motion for
delayed two time Green functions
\begin{equation}
\omega\ll A|B \gg_\omega=\langle[A, B]\rangle+\ll[A, H] | B\gg \label{A1}
\end{equation}
where
\begin{equation}
\ll A|B \gg_\omega=-i\int\limits_0^\infty dt \exp\left[i(\omega+i\varepsilon)
t\right]\left\langle\left[A(t), B(0)\right]\right\rangle\left.
\right|_{\varepsilon\to+0} \label{A2}
\end{equation}
The bracketed expression indicates the commutator of operators, and that within
the French quats --- the averaging over the Gibbs ensemble. For the Hamiltonian
(\ref{4}-\ref{7}) we have the accurate equation:
\widetext
\begin{eqnarray}
(\omega-\omega_\lambda)\ll b_\lambda | b_\lambda^+ \gg_\omega & =&1+3
\sum_{\lambda_2 \lambda_3}
\Phi^{(3)}(-\lambda, \lambda_2, \lambda_3) \ll A_{\lambda_2} A_{\lambda_3} |
b_\lambda^+ \gg_\omega\nonumber \\
&&+4\sum_{\lambda_2 \lambda_3 \lambda_4} \Phi^{(4)}
(-\lambda, \lambda_2, \lambda_3, \lambda_4)
\ll A_{\lambda_2} A_{\lambda_3} A_{\lambda_4} |
b_\lambda^+ \gg_\omega \label{A3}
\end{eqnarray}
With an accuracy up to
$\varkappa^2$ the Green function describing the contribution of
four-phonon processes to (\ref{A3}) can be decoupled
\begin{eqnarray}
\ll A_{\lambda_2} A_{\lambda_3} A_{\lambda_4} | b_\lambda^+ \gg_\omega&=&
\langle A_{\lambda_2} A_{\lambda_3}\rangle
\ll A_{\lambda_4} | b_\lambda^+ \gg_\omega+ \langle A_{\lambda_2}
A_{\lambda_4}\rangle \ll A_{\lambda_3} | b_\lambda^+ \gg_\omega\nonumber \\
&&+\langle A_{\lambda_3} A_{\lambda_4}\rangle \ll A_{\lambda_2} | b_\lambda^+ \gg_\omega
\nonumber\\
&=&
\delta_{\lambda_2 \lambda_3}(1+2N_{\lambda_2})\delta_{\lambda_4\lambda}+
\delta_{\lambda_2
\lambda_4}(1+2N_{\lambda_1})\delta_{\lambda_3\lambda}\nonumber \\
&&+\delta_{\lambda_2 \lambda_4}(1+2N_{\lambda_3})\delta_{\lambda_2\lambda}+
\ll b_\lambda | b_\lambda^+ \gg_\omega \label{A4}
\end{eqnarray}
This is immediately
followed by the result (\ref{12}) for the contribution of four-phonon processes to
the frequency shift. For the determination of the contribution of three-phonon
processes we must write the chain of equations
for the Green-functions entering the following relation
\begin{eqnarray}
\ll A_{\lambda_2} A_{\lambda_3} | b_\lambda^+ \gg_\omega &=&\ll b_{\lambda_2} b_{\lambda_3} |
b_\lambda^+ \gg_\omega +\ll b_{-\lambda_2} b_{\lambda_3} |
b_\lambda^+ \gg_\omega\nonumber \\
&&+\ll b_{\lambda_2} b_{-\lambda_3} |
b_\lambda^+ \gg_\omega +\ll b_{-\lambda_2} b_{-\lambda_3} |
b_\lambda^+ \gg_\omega  \label{A5}
\end{eqnarray}
\narrowtext
and make there decoupling. Upon separation of the real and
imaginary parts of Green function
$\ll b_\lambda | b_\lambda^+ \gg_\omega$ 
we obtain immediately the result
(\ref{9}-\ref{13}). The quasiharmonic frequency shift proves to be then as expressed
through the amplitudes of three-phonon processes:
\begin{equation}
\Delta_\lambda^{(qh)}=-36\sum_{\lambda_2 \lambda_3} \frac{1+2N_{\lambda_3}}
{\omega_{\lambda_2}}
\Phi(\lambda,-\lambda, \lambda_2)\Phi(\lambda_3,-\lambda_3,-\lambda_2) \label{A6}
\end{equation}
As far as we know this
expression is new, although the relation itself of thermal expansion to
three-phonon anharmonism was qualitatively discussed as early as in the
classical book by Paierls \cite{A2}.

\begin{figure}
\caption{Gruneisen parameters for fcc phase of Ca (solid line)
and Sr (dashed line).}
\label{fig3}
\end{figure}

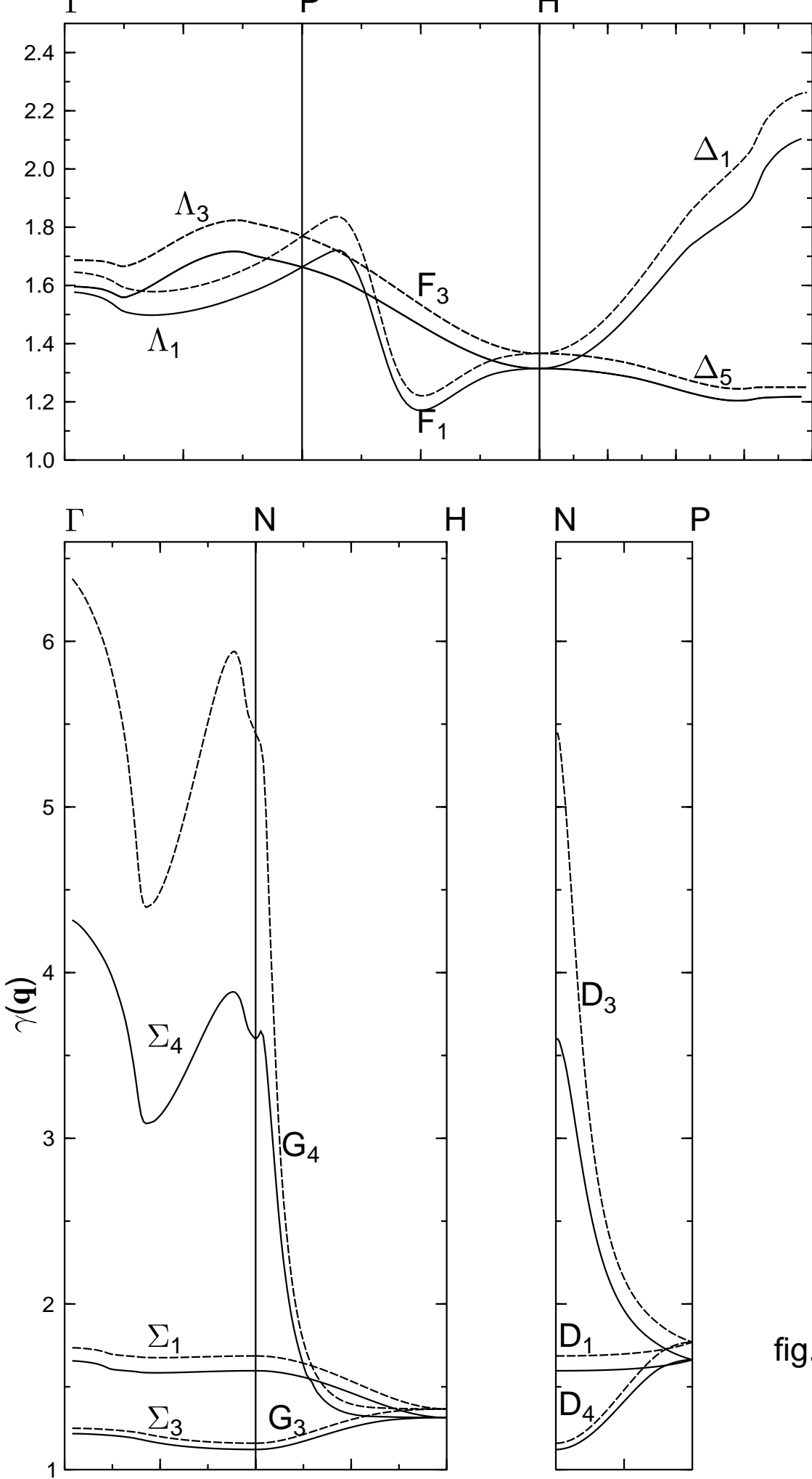
\begin{figure}
\caption{Gruneisen parameters for bcc phase of Ca (solid line) and Sr
(dashed line).}
\label{fig4}
\end{figure}

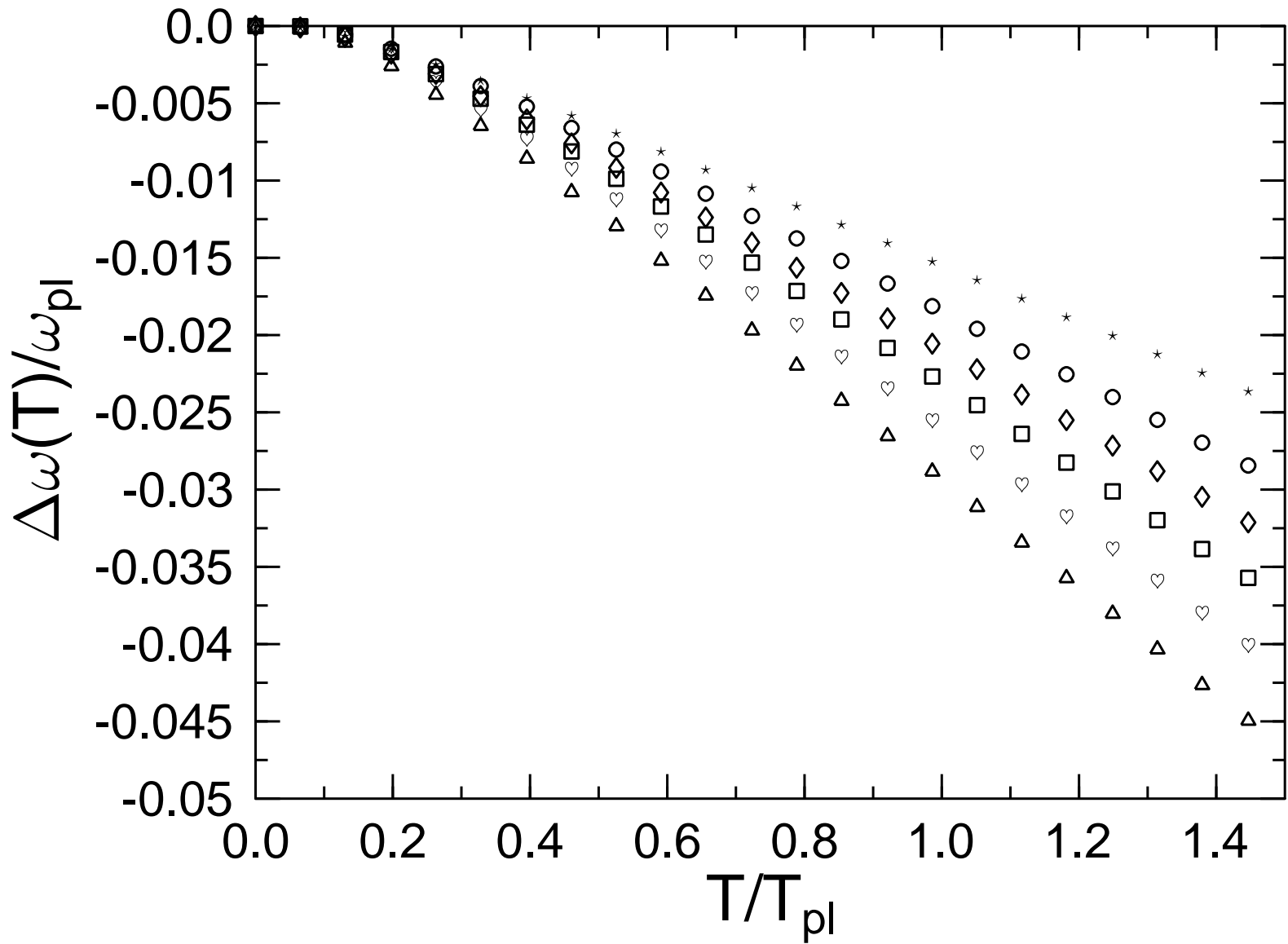
\begin{figure}
\caption{Temperature dependence of frequency shift determined by Eq.
(\ref{extra1}) at symmetric points of the
Brillouin zone for fcc Ca;
$\omega_{pl}=(4\pi Ze^2/M\Omega)^{1/2}$
is the plasma ion frequency, $T_{pl}$ is the corresponding
temperature (for Ca $T_{pl}$=476 K.)  The symbols correspond
to different vibration modes: $\circ$ --- $W_1$,
$\triangle$ --- $W_2$, $\square$ --- $X_1$,
$\diamondsuit$ --- $X_2$,
$\heartsuit$ --- $L_1$,  $\star$ --- $L_2$.}
\label{fig5a}
\end{figure}

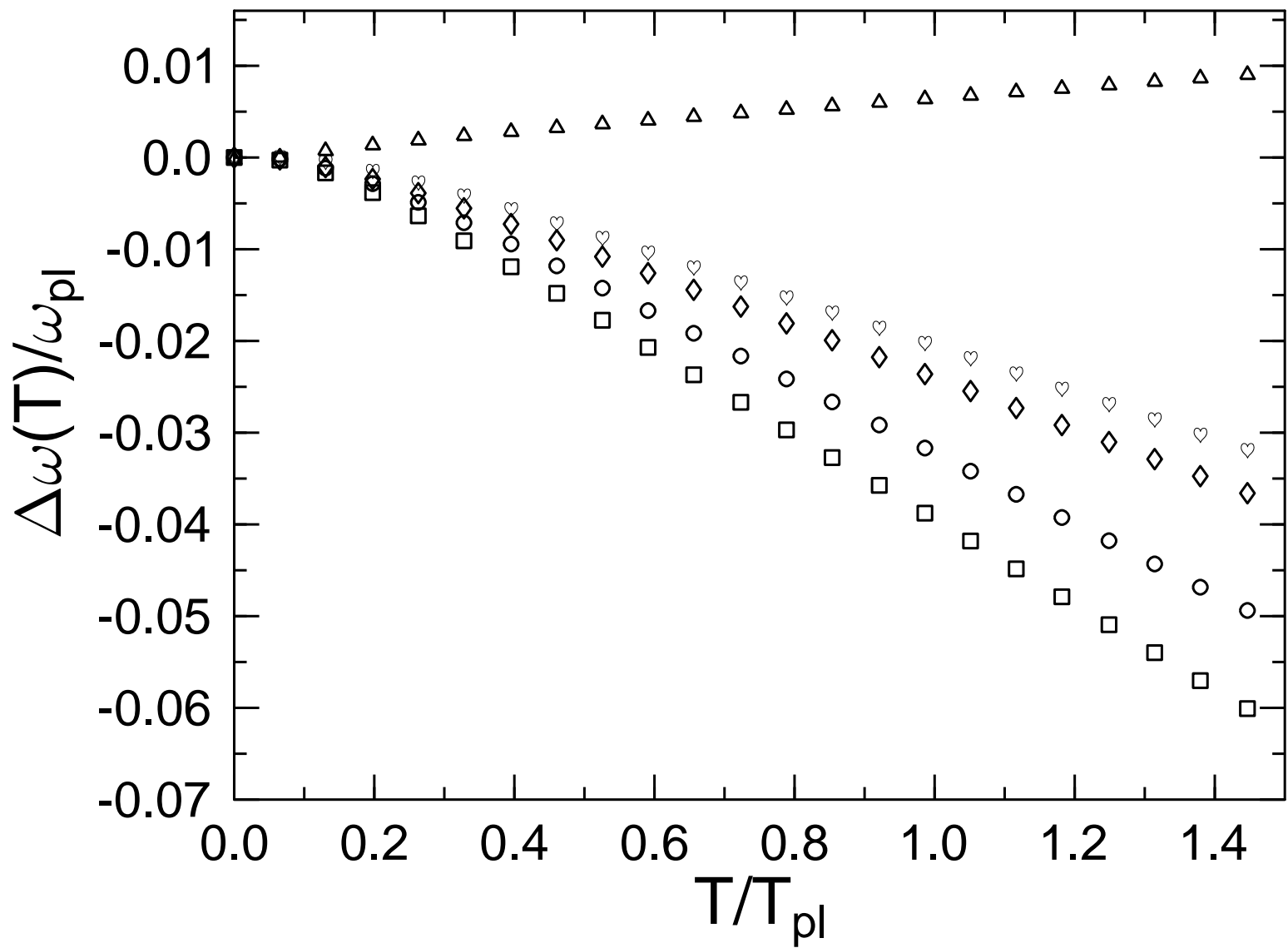
\begin{figure}
\caption{Temperature dependence of frequency shift at symmetric points of the
Brillouin zone for bcc Ca;  The symbols correspond
to different vibration modes: $\circ$ --- $N_1$, $\diamondsuit$ --- $N_3$
$\triangle$ --- $N_4$, $\square$ --- $H$,
$\heartsuit$ --- $P$.}
\label{fig5b}
\end{figure}

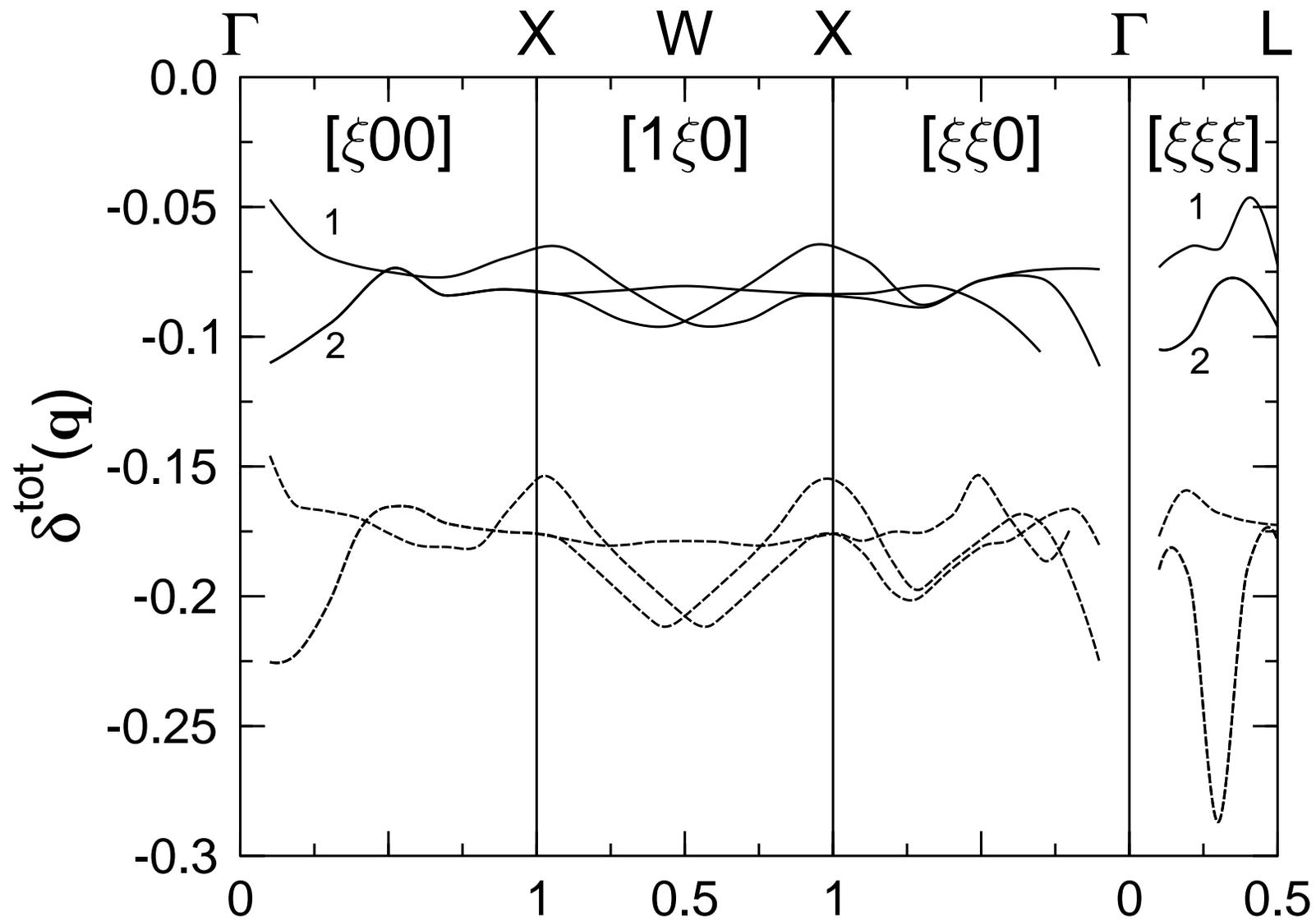
\begin{figure}
\caption{Change of the relative frequency shift
$\delta_{\xi\mathbf{k}}(T)=
\Delta\omega_{\xi\mathbf{k}}(T)/\omega_{\xi\mathbf{k}}(0)$
in the Brillouin zone
for fcc phases of Ca and Sr at the temperature of structure transition $T=T_s$
($T_s=726$ K for
Ca and 930 K for Sr), $\Delta\omega_{\xi\mathbf{k}}(T)/$ is determined by
Eq. (\ref{extra1}).
Solid line --- Ca, dashed
line --- Sr.}
\label{fig6a}
\end{figure}

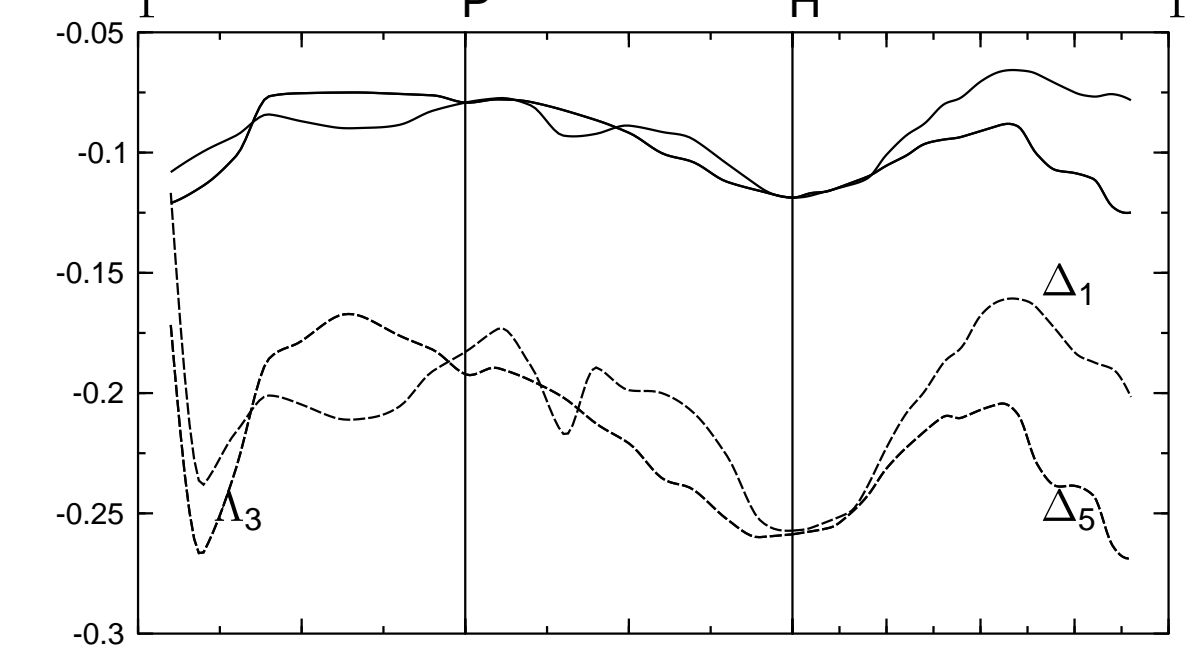
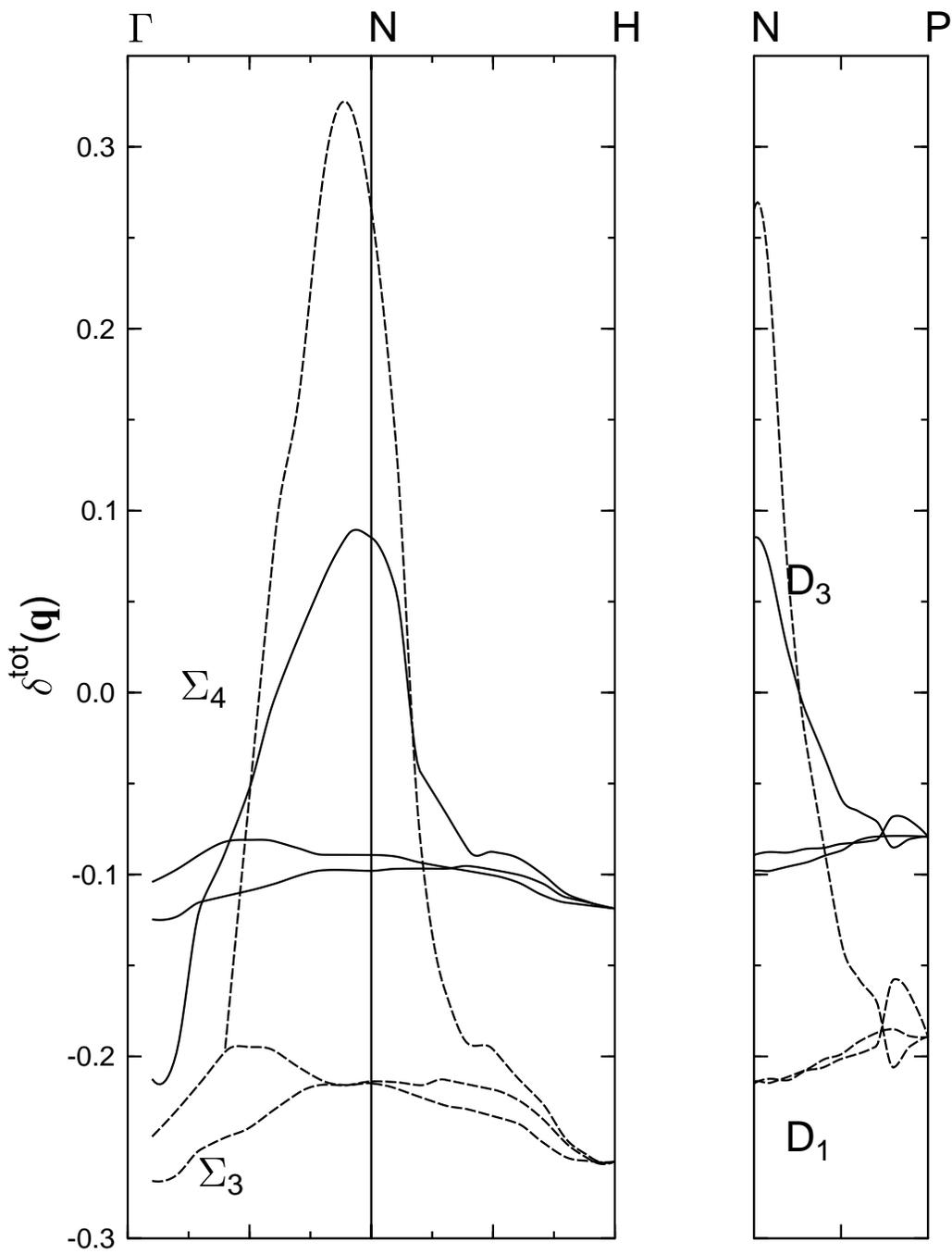
\begin{figure}
\caption{The same as Fig. \ref{fig6a} for bcc phase of Ca and Sr.}
\label{fig6b}
\end{figure}

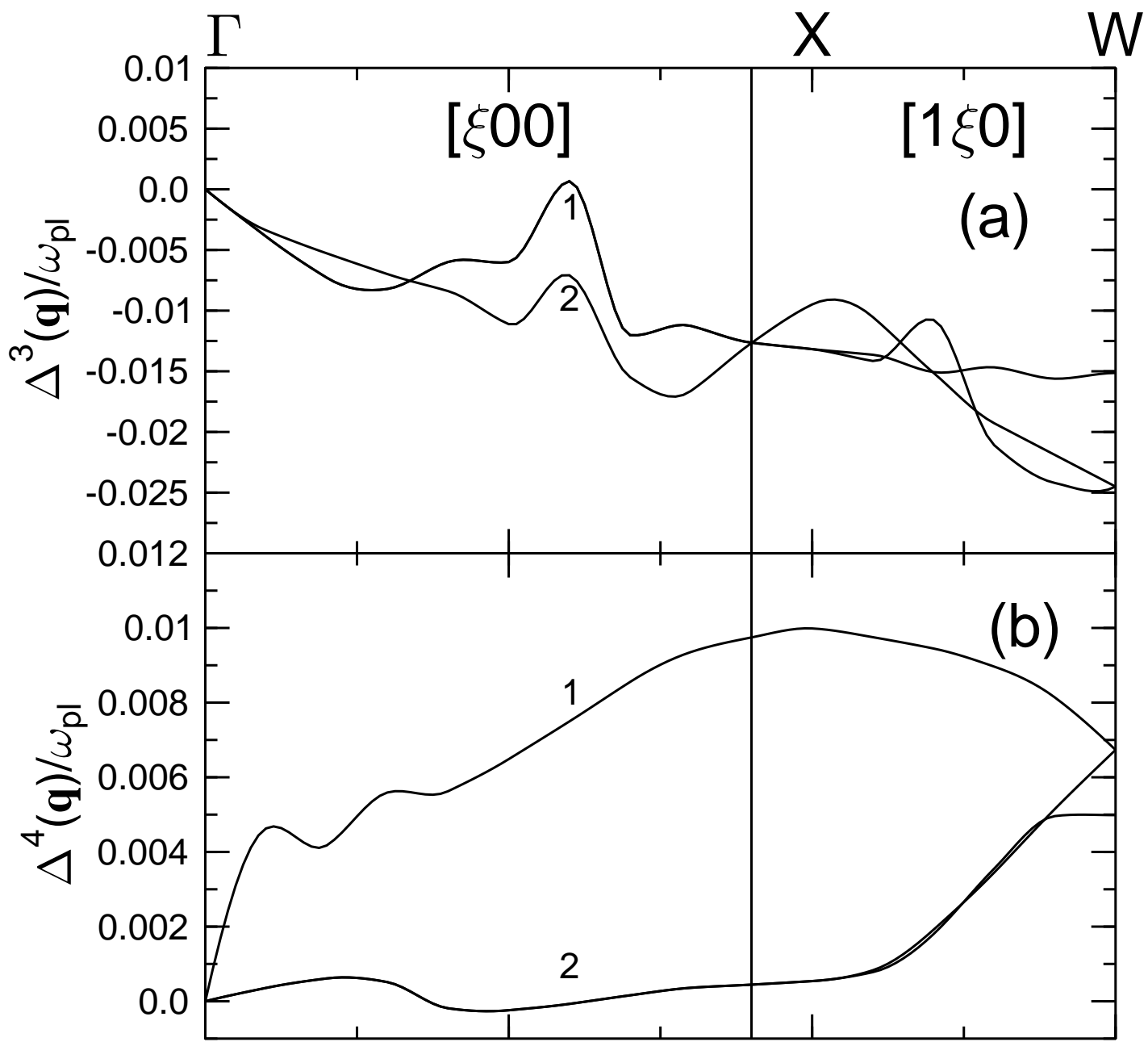
\begin{figure}
\caption{Change in the Brillouin zone of the anharmonic frequency shift determined by
formulae (\ref{11}), (\ref{12}) for fcc Ca at $T=T_s$.}
\label{fig7a}
\end{figure}

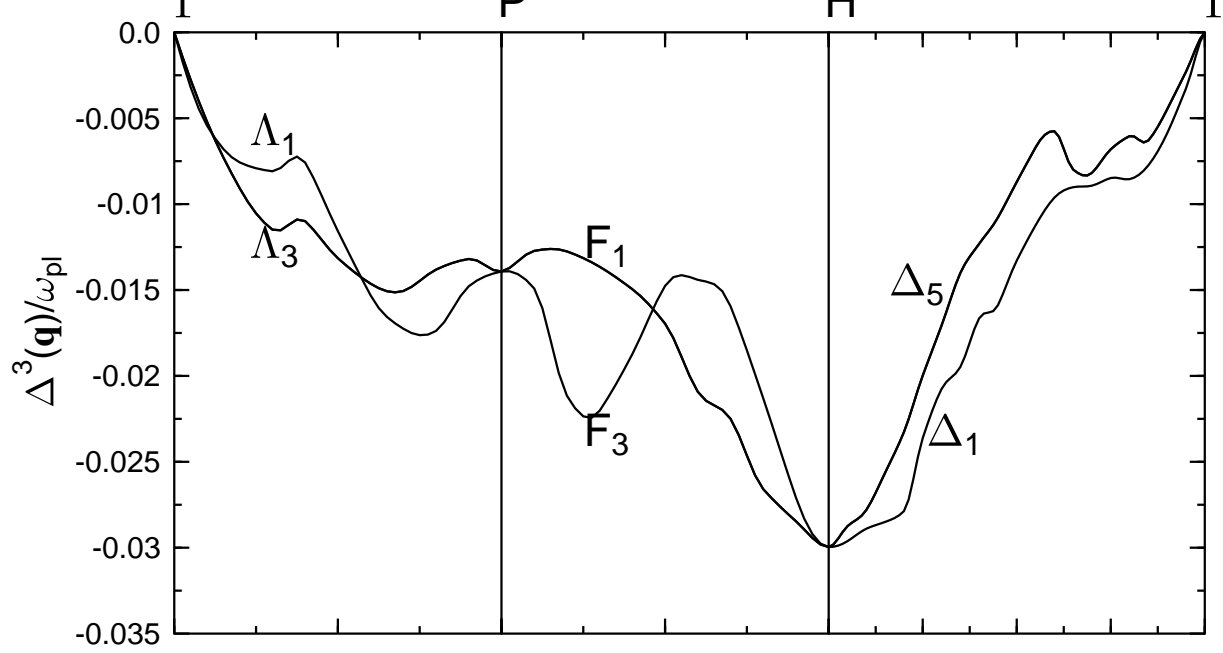
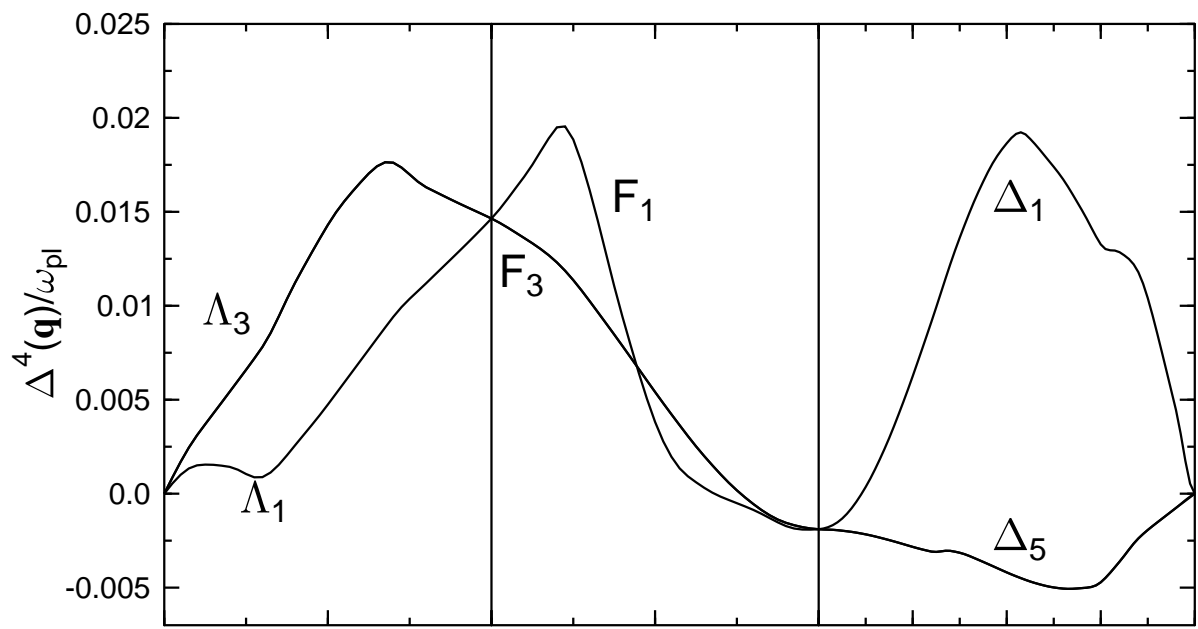
\begin{figure}
\caption{The same as Fig. \ref{fig7a} for bcc phase of Ca.}
\label{fig7b}
\end{figure}

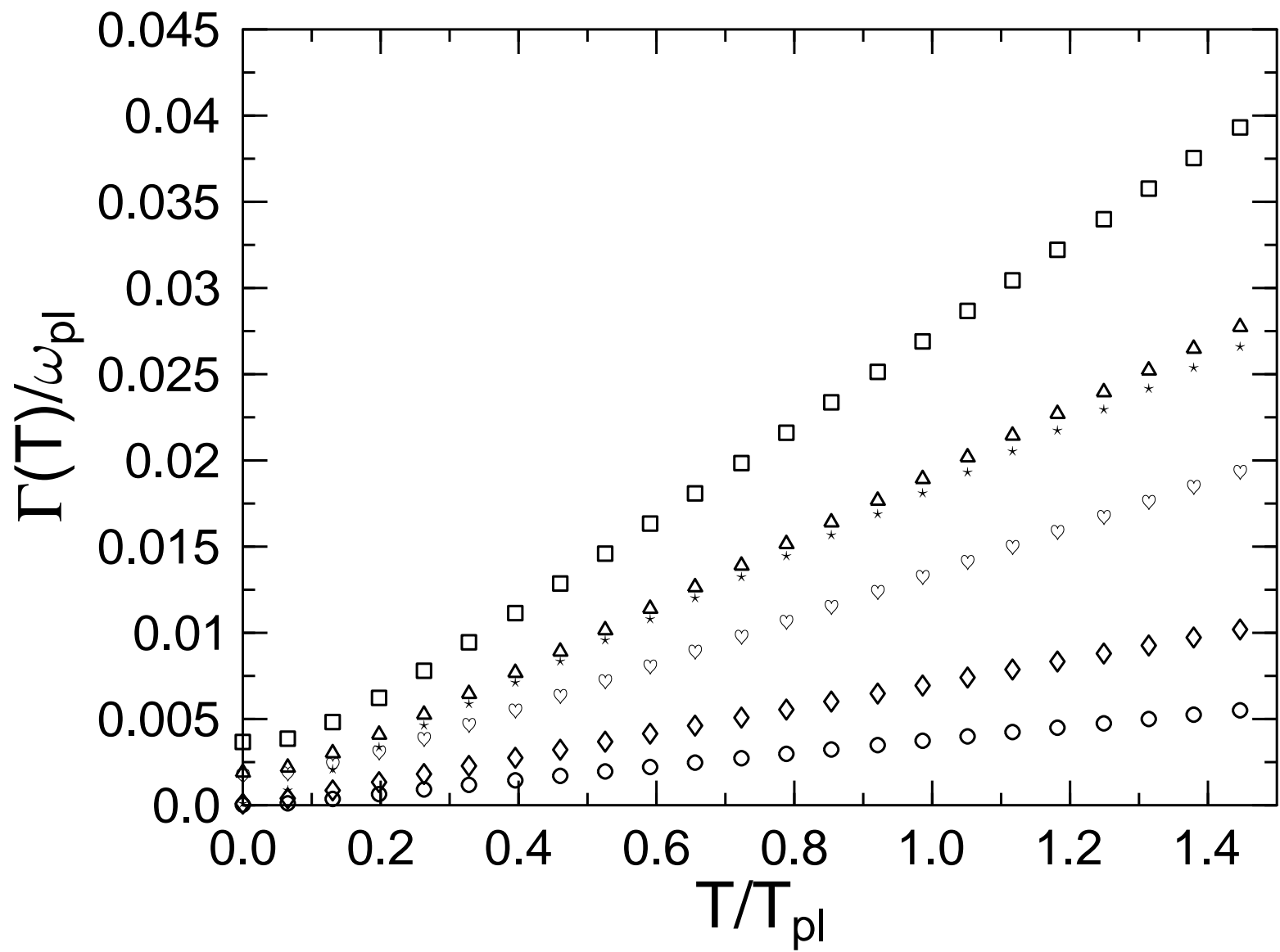
\begin{figure}
\caption{Temperature dependence of phonon damping at symmetric  points of
Brillouin zone for fcc Ca. $T_{pl}$ and the symbols are
the same as in Fig. \ref{fig5a}.}
\label{fig8a}
\end{figure}

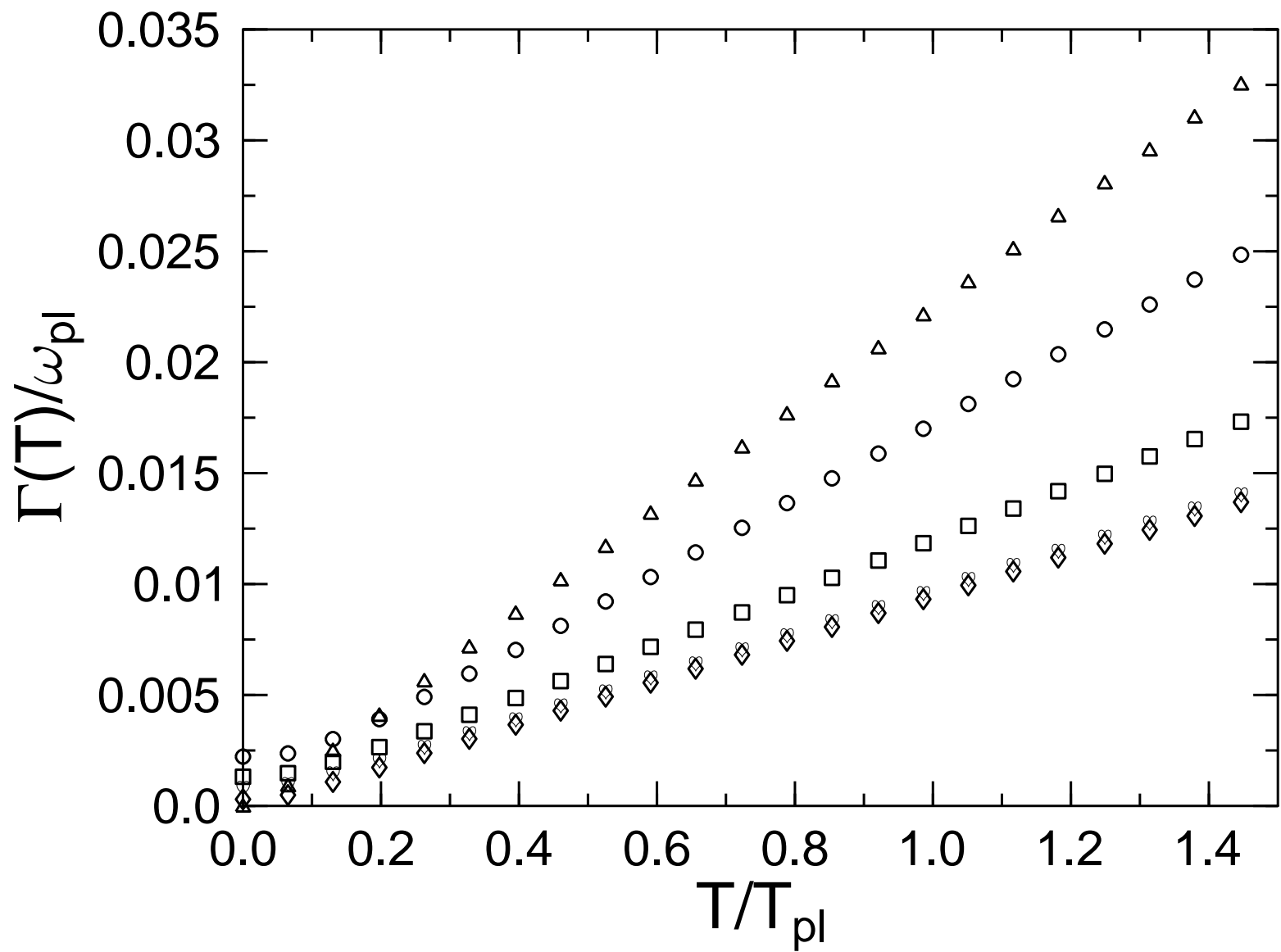
\begin{figure}
\caption{The same as Fig. \ref{fig8a} for bcc phase of Ca. The symbols are
the same as in Fig. \ref{fig5b}.}
\label{fig8b}
\end{figure}

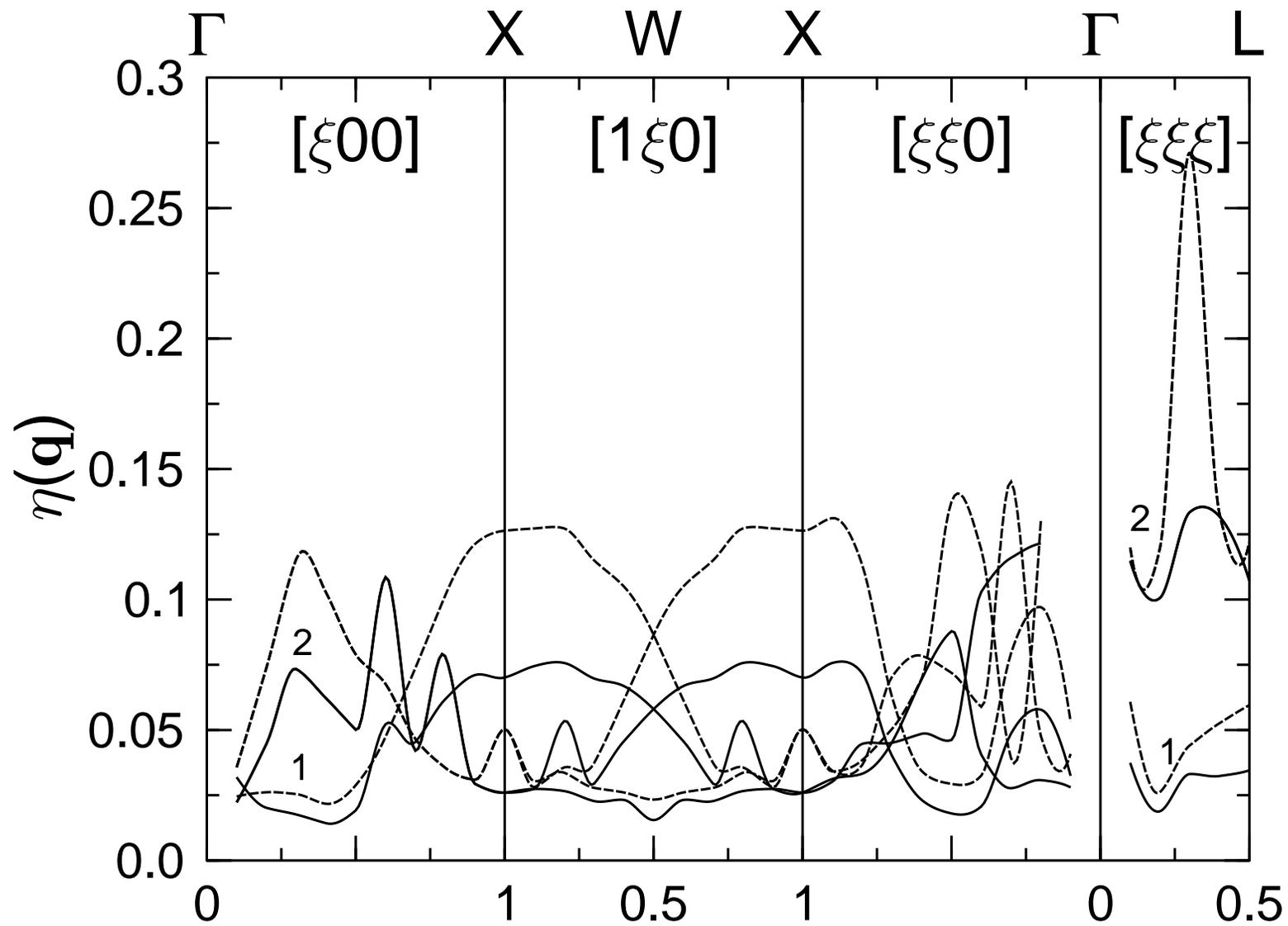
\begin{figure}
\caption{Change in the Brillouin zone of relative damping
$\eta_{\xi\mathbf{k}}(T)=\Gamma_{\xi\mathbf{k}}(T)/\omega_{\xi\mathbf{k}}(0)$
for fcc Ca and Sr at
$T=T_s$. Solid line --- Ca, dashed line --- Sr.}
\label{fig9a}
\end{figure}

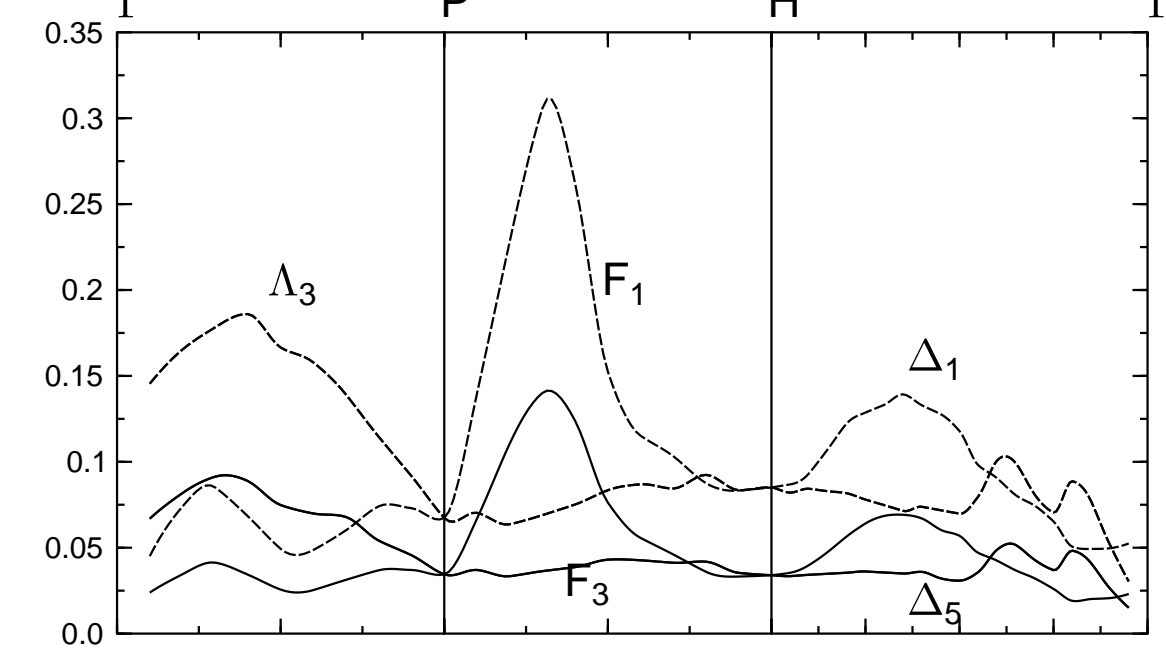
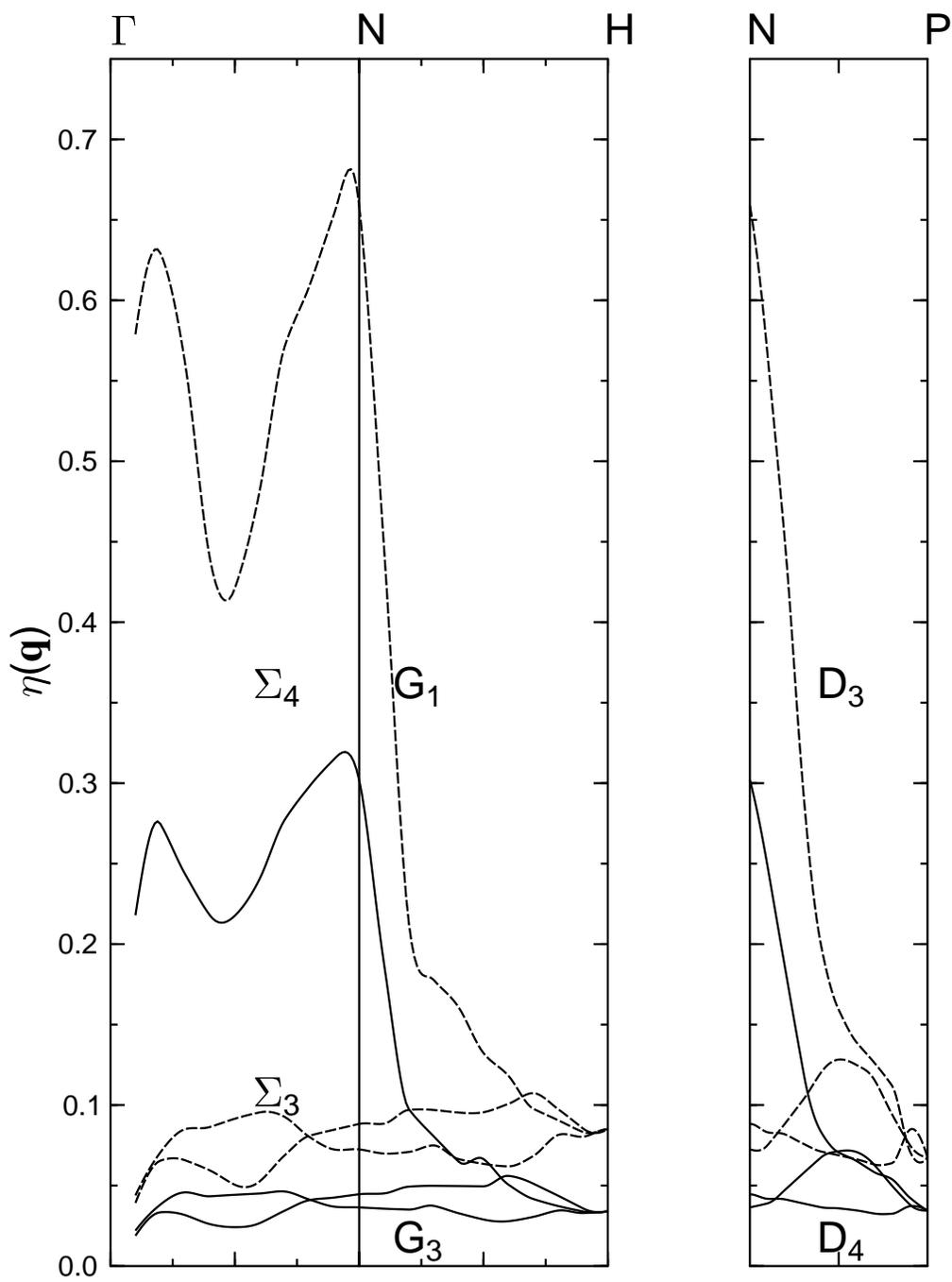
\begin{figure}
\caption{The same as Fig. \ref{fig9a} for bcc phase of Ca and Sr.}
\label{fig9b}
\end{figure}

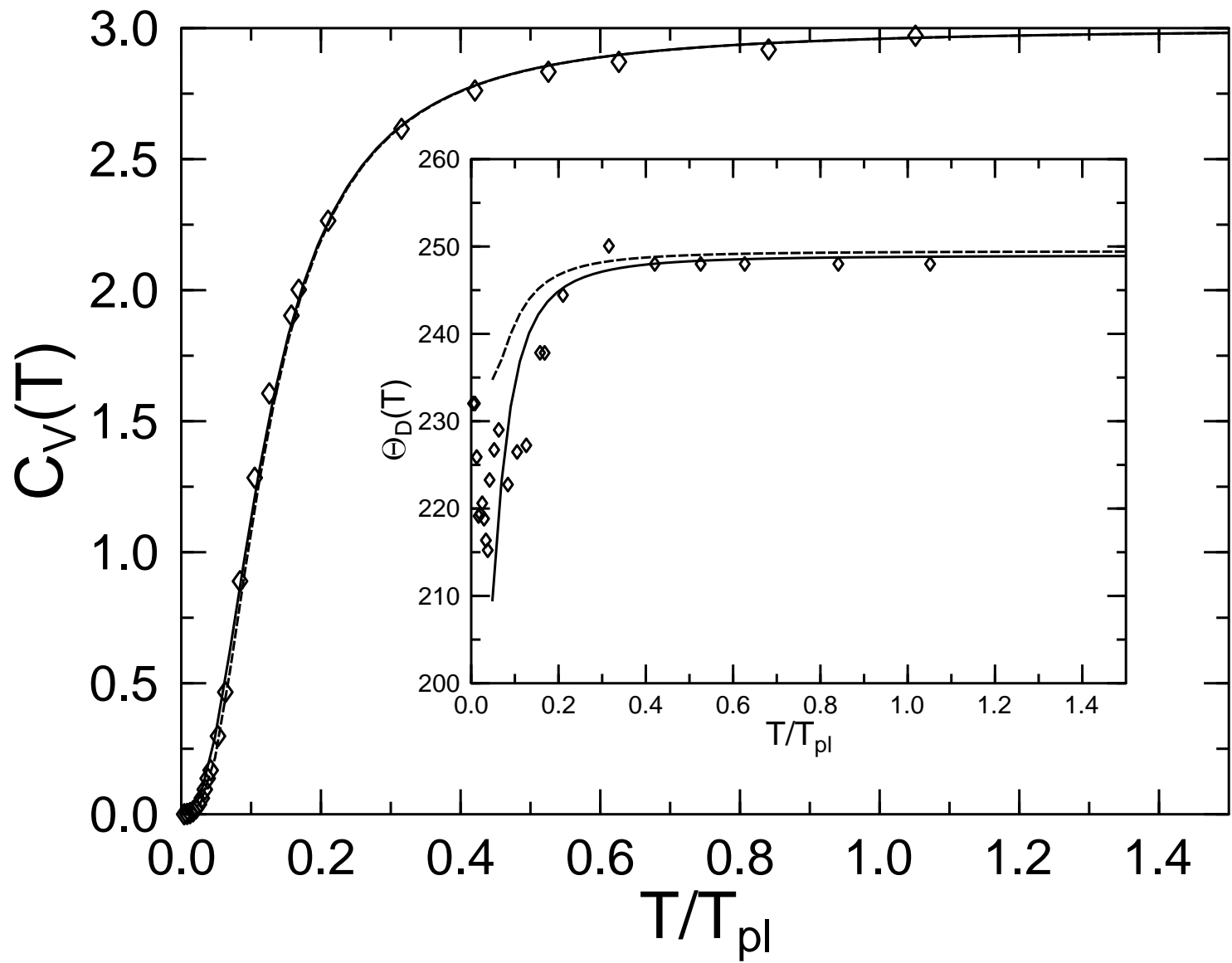
\begin{figure}
\caption{Phonon contribution to the lattice heat capacity of Ca;
in the insert the temperature dependence of effective Debye temperature is
shown.  $\diamondsuit$ is the result of experimental treatment
(see the text); solid line --- the
calculation result for the bcc phase, dashed line --- for the fcc phase.}
\label{fig10a}
\end{figure}

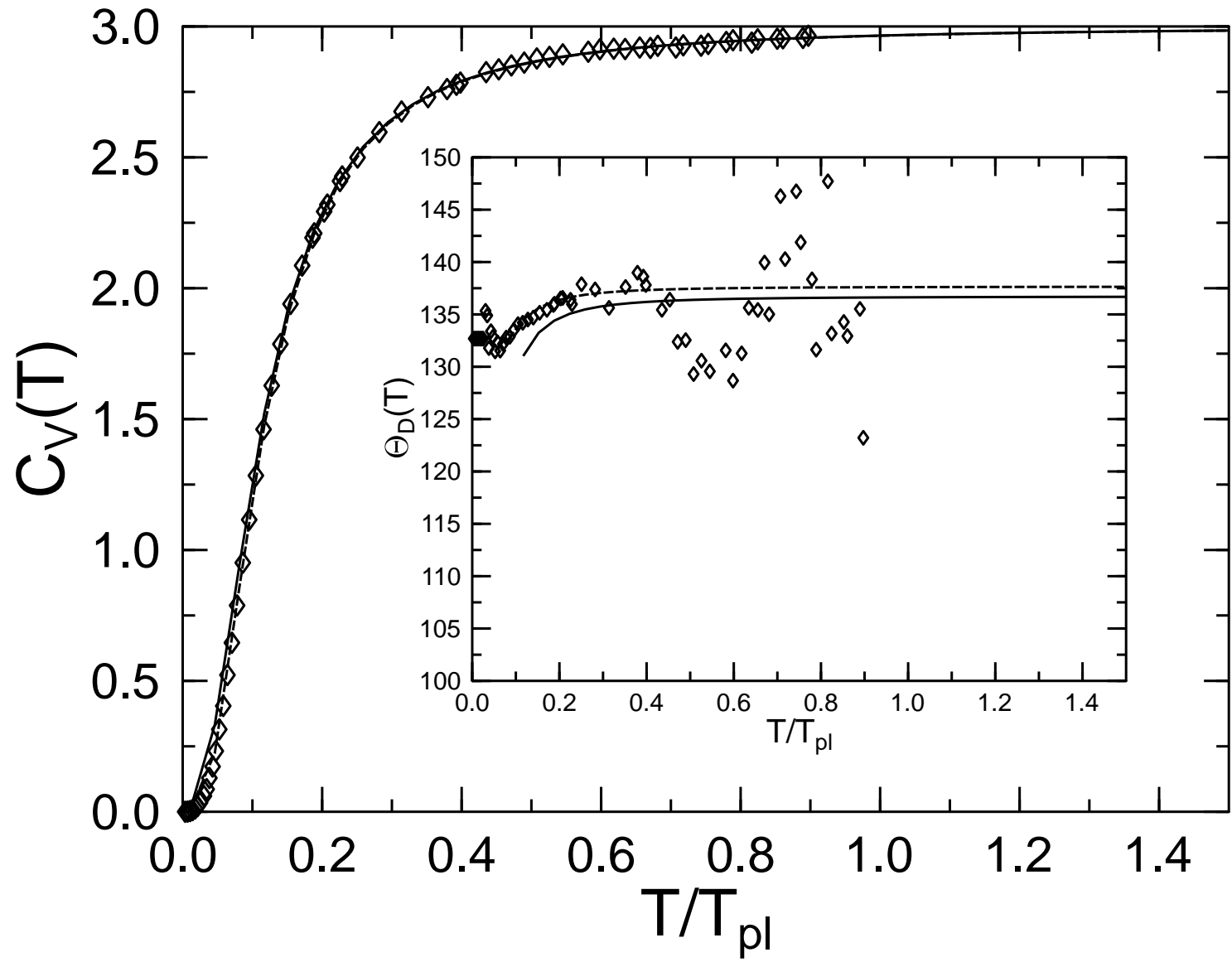
\begin{figure}
\caption{The same as Fig. \ref{fig10a} for Sr.}
\label{fig10b}
\end{figure}

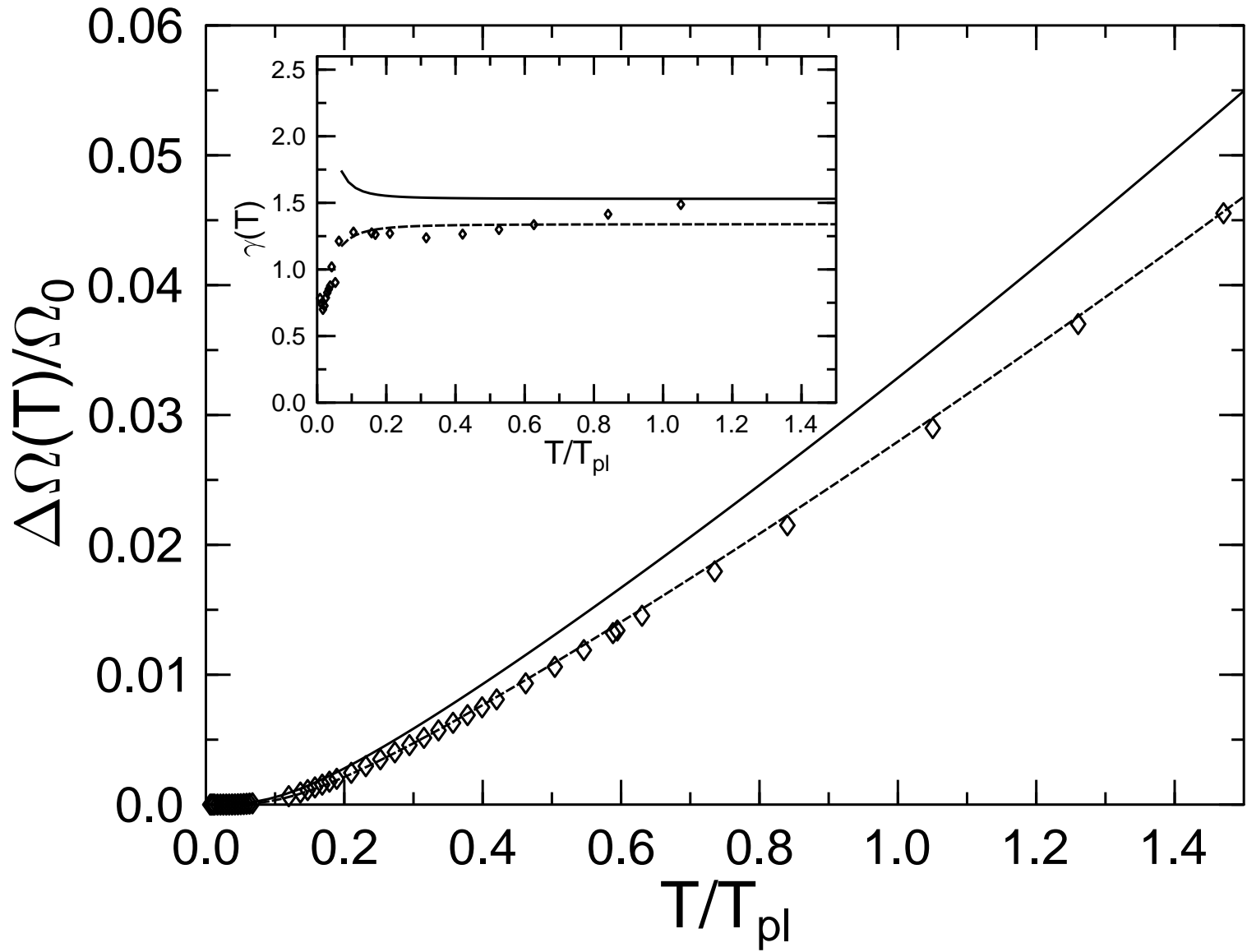
\begin{figure}
\caption{Phonon contribution to the thermal expansion of Ca; in the
insert the temperature dependence of  Gruneisen parameter is shown. The meaning
of symbols and lines is the same as in Fig. \ref{fig10a}.}
\label{fig11a}
\end{figure}

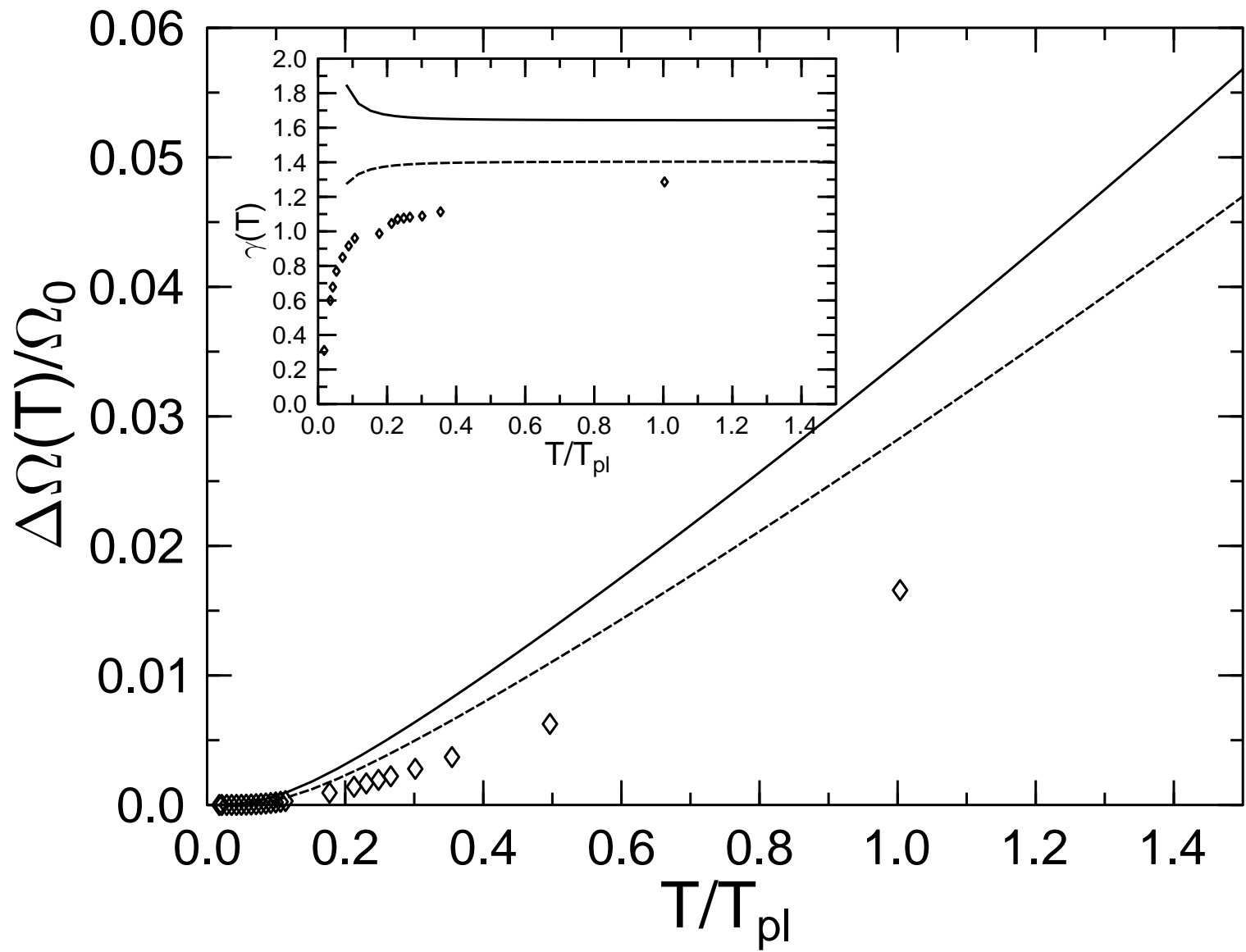
\begin{figure}
\caption{The same as Fig. \ref{fig11a} for Sr.}
\label{fig11b}
\end{figure}

\mediumtext
\begin{table}
\caption{Model parameters.}
\begin{tabular}{ccddcddc}
Metal & $T_s$ & $\Omega_{exp} (au)$ & $\omega_{pl}$ (THz)&
$T_m$ (K)& $r_0$ (au) & U & Z\\
\tableline
Ca & 726 & 303.4 & 9.91 & 1112 & 2.545 & 0.636 & 2\\
Sr & 930 & 394.6 & 5.88 & 1045 &3.024 & 0.787 & 2\\
\end{tabular}
\label{table1}
\end{table}

\begin{table}
\caption{Phonon spectra momenta.}
\begin{tabular}{cdddddd}
momenta & \multicolumn{3}{c}{Ca}&\multicolumn{3}{c}{Sr}\\
& experiment & theory & theory & experiment &
theory & theory \\ & FCC & FCC & BCC & FCC & FCC & BCC \\
\tableline
$\Omega_{log}$, K & 136 & 138 & 128 & 81  & 78  & 67 \\
$\Omega_{-2}$, K  & 130 & 142 & 121 & 77  & 77  & 64 \\
$\Omega_{-1}$, K  & 152 & 158 & 148 & 89  & 88  & 80 \\
$\Omega_1$, K     & 181 & 185 & 182 & 102 & 102 & 99 \\
$\Omega_2$, K     & 195 & 193 & 193 & 107 & 107 & 106 \\
$\Omega_4$, K     & 214 & 207 & 208 & 116 & 114 & 114 \\
$\Omega_*$, K      & 220 &     &     & 120 &     &     \\
$\gamma$, $\frac{mJ}{mol\, K^2}$ & 3&&&13.8&&\\
$\beta$, $\frac{mJ}{mol\, K^4}$& 1.87$\times 10^{-2}$&&&8.9
$\times 10^{-2}$ &&\\
$\Theta_{LT}$, K&232&&&138&&\\
A, $\frac{mJ}{mol\, K^2}$&9.1&&&6.9&&\\
$\Theta_{HT}, K$ & 252 & 249 & 249 & 139 & 138 & 137 \\
\end{tabular}
\label{table2}
\end{table}

\begin{table}
\caption{Temperature dependence of  Gruneisen parameter.}
\begin{tabular}{cddddd}
                  & $N_1$ & $N_3$ & $N_4$& H     & P   \\
\tableline
\multicolumn{6}{c}{Ca, BCC T=726 K} \\
$\gamma_0$        & 1.60  & 1.12  & 3.60  & 1.31  & 1.66 \\
$\Delta\gamma(T)$ & 0.059 & 0.12  & -2.87 & 0.24  & -0.13 \\
$\gamma(T)$       & 1.66  & 1.24  & 0.73  & 1.55  & 1.53 \\
\multicolumn{6}{c}{Sr, BCC T=930 K} \\
$\gamma_0$        &1.69   & 1.16  & 5.44  & 1.37  & 1.77 \\
$\Delta\gamma(T)$ &0.13   & 0.31  & -20.0 & 0.33  & -0.37 \\
$\gamma(T)$      & 1.82  & 1.47  & -14.6 & 1.7   & 1.40 \\
\end{tabular}
\label{table3}
\end{table}
\end{document}